\documentclass[preprint,aps,pre,groupedaddress,showpacs,preprintnumbers,amsmath,amssymb]{revtex4}
\usepackage{bm,amsmath,amsthm}
\usepackage{graphicx}
\usepackage{subfigure}
\usepackage{dcolumn}

\newcommand\eq[1]{\begin{eqnarray}#1\end{eqnarray}}

\newcommand{\inte}{\!\int\!\!}
\def\dbar{{\mathchar'26\mkern-12mu d}}
\newcommand{\ud}{\dbar}

\newcommand{\D}{\mathcal D}

\newcommand{\mar}{\bm r}
\newcommand{\maxx}{\bm x}

\newcommand{\maq}{\bm q}
\newcommand{\map}{\bm p}

\begin{document}
\title{Anomalous scaling and anisotropy in models of passively advected vector fields}
\author{Heikki Arponen}
\affiliation{Helsinki University, Department of Mathematics and
Statistics, P.O. Box 68, 00014 Helsinki (Finland)}
\email{heikki.arponen@helsinki.fi}
%
%
%
%
%
%
\date{\today}
\begin{abstract}
An anisotropically forced passive vector model is analyzed at
scales much smaller and larger than the forcing scale by solving
exactly the equation for the pair correlation function. The model
covers the cases of magnetohydrodynamic turbulence, the linear
pressure model and the linearized Navier-Stokes equations by
choice of a simple parameter. We determine whether or not the
anisotropic injection mechanism induces dominance of the
anisotropic effects at the asymptotic scaling regimes. We also
show that under very broad conditions, both scaling regimes
exhibit anomalous scaling due to the existence of nontrivial zero
modes.
\end{abstract}
\maketitle
%
%
%
\section{Introduction}
One of the most important problems of turbulence is the observed
deviation of Kolmogorov scaling in the structure functions of the
randomly stirred Navier-Stokes equations in the inertial range of
scales \cite{frisch}. Contemporary research in turbulence has
recently provided an explanation for this phenomenon in the
context of passive advection models (see e.g. \cite{falkovich} for
an introduction and further references). In the case of the
passive scalar model describing the behavior of a dye
concentration in a turbulent fluid, such a violation of canonical
scaling behavior (henceforth referred to as anomalous scaling) has
recently been traced to the existence of a type of statistical
integrals of motion known as zero modes
\cite{falkovich,gawedzkikupiainen}. The result can be obtained
under some simplifying assumptions about the velocity field,
namely assuming the velocity statistics to be gaussian and white
noise in time, which results in a solvable hierarchy of Hopf
equations for the correlation functions. Such properties are
included in the so called Kraichnan model \cite{kraichnan} of
velocity statistics, which will also be utilized in the present
work.\\ \\
As opposed to a thermodynamical equilibrium, the passive scalar is
maintained in a nonequilibrium steady state by external forcing
designed to counter molecular diffusion. It was proved in
\cite{ville} that even in the limit of vanishing molecular
diffusivity the steady state exists and is unique. Furthermore
defining the integral scale to be infinity results in an infinite
inertial range, divided only by the injection scale $L$ due to the
forcing. While the above results of the passive scalar anomalous
scaling were concerned with the small scale problem $r \ll L$, in
\cite{fouxon} it was observed that one obtains anomalous scaling
also at large scales, provided the forcing is of "zero charge",
$q_0 \doteq \int \ud^d \mar C_L (\mar) =0$, where $C_L$ is the
forcing pair correlation function. Such a forcing is concentrated
around finite wavenumbers $k \sim 1/L$, which behaves similarly to
a zero wavenumber concentrated forcing at small scales, but is
more realistic for probing scales larger than the forcing scale.\\ \\
The forcing is usually taken to be statistically isotropic.
Justification for this is that one usually expects the anisotropic
effects to be lost anyway at scales much smaller that the forcing
scale, according to a universality hypothesis by the K41
theory\cite{frisch}. Nevertheless, in \cite{celani} it was
discovered that even a small amount of anisotropy in the forcing
(that can never be avoided in a realistic setting) in the passive
scalar equation would render the large scale behavior to be
dominated by anisotropic zero modes responsible for another type
of anomalous scaling. As pointed out in \cite{celani}, such
behavior is nontrivial also in the sense that one might expect the
system to obey Gibbs statistics with exponentially decaying
correlations at large scales, as indeed happens for the pair
correlation function with isotropic zero charge forcing
\cite{fouxon}.
\\ \\
The purpose of the present work is to consider the small and large
scale behavior of passive \emph{vector} models stirred by an
anisotropic forcing, and especially to determine if the phenomena
of anomalous scaling and persistence of anisotopy is a general
feature of passive advection models or just a curiosity of the
passive scalar. The passive vector models arise as quite natural generalizations of the scalar problem and turn out to possess much richer phenomena already at the level of the pair correlation function. For example the pair correlation function of the magnetohydrodynamic equations exhibit anomalous scaling \cite{vergassola} whereas one needs to study the fourth and higher order structure functions of the passive scalar to see such behavior (see e.g. \cite{paolo.antti} and references therein). It has also been argued that the linear passive vector models might yield the exact scaling exponents of the full Navier-Stokes turbulence \cite{angheluta}. The equation under study is defined as
\eq{ \dot u_i - \nu \Delta u_i + \mathbf v  \cdot \nabla u_i - a
\mathbf u \cdot \nabla v_i  + \nabla_i P = f_i ,\label{equation}}
with a parameter $a = -1,0$ or $1$, corresponding respectively to
the linearized Navier-Stokes equations (abbreviated henceforth as
LNS), the so called linear pressure model (LPM) and the
magnetohydrodynamic (MHD) equations. $\nu$ is a constant
viscosity/diffusivity term, $f_i$ denotes an external stirring
force, $v_i$ is a gaussian, isotropic external velocity field
defined by the Kraichnan model and $P$ is the pressure, giving
rise to nonlocal interactions. The equation was introduced in
\cite{paolo}, where the authors derived and studied a zero mode
equation for the pair correlation function in the isotropic sector
and found the small scale exponents numerically and to a few first
orders in perturbation theory (see also \cite{jurc1} for a more
detailed exposition). They also reported perturbative results for
higher order correlation functions and anisotropic sectors using
the renormalization group. Although the purpose of the present
work is to consider arbitrary values of $a$, some cases have
already been studied elsewhere. The $a=1$ case, corresponding to
magnetohydrodynamic turbulence, has probably received the most
attention\cite{antonovlanottemazzino,vergassola,arponen,arad,jurc2,lanotte2}.
The linear pressure model (or just the passive vector model) with
$a=0$, has been studied in e.g.\cite{jurc1,adz,benzi}. The
linearized Navier-Stokes equation (see \cite{frisch}), $a=-1$, was
studied in \cite{jurc1} and numerically in \cite{yoshida} in two
dimensions and is the least known of the above cases, although
perhaps the most interesting. The above mentioned studies have
been restricted to the small scale problem and rely heavily on the
zero mode analysis, i.e. finding the homogeneous solutions to the
pair correlation equation. For our purposes this is not enough. To
capture the anomalous properties as discussed above, one needs to
consider the amplitudes of the zero modes as well, as it may turn
out that some amplitudes vanish. Indeed, it is exactly this sort
of mechanism that is responsible for the
anisotropy dominance in \cite{celani}. \\ \\
We provide an exact solution of the equation for the pair
correlation function with anisotropic forcing and study both small
and large scale behavior. It turns out that for the "zero charge"
forcing as above, the large scale behavior is anomalous even in
the isotropic sector for all $a$. The anisotropy dominance seems
however rather an exception than a rule in three dimensions, as
only the trace of the correlation function for the $a=0$ model
exhibits similar phenomena at large scales. Nevertheless, in two
dimensions the anisotropy dominance is a more common phenomenon.
Perhaps the most interesting case is the linearized Navier-Stokes
equation for which $a=-1$. The field $u$ is now considered to be a
small perturbation to the steady turbulent state described by $v$.
This case is unfortunately complicated by the fact that practically
nothing is known of the existence of the steady state, although an
attempt to rectify
the situation is underway by the present author.\\ \\
In section \ref{setup} we introduce the necessary tools, discuss
the role of the forcing and present the equation for the pair
correlation function in a Mellin transformed form. Details of it's
derivation can be found in appendix \ref{APXeom}. In section
\ref{solution} we present the solution in both isotropic and
anisotropic sectors and explain the results for the passive scalar
of \cite{celani} in our formalism. The next three sections are
concerned with the specific cases of magnetohydrodynamic
turbulence, linear pressure model and the linearized Navier-Stokes
equations. Although the space dimension is arbitrary (although
larger than or equal to two), we concentrate mostly on two and
three dimensions. The reasons for this are the considerable
differences between $d=2$ and $d=3$ cases and the similarities of
dimensions $d \ge 3$. Mainly one may expect some sort of
logarithmic behavior in two dimensions while in higher dimensions
the behavior is power law like. Also the presence of anomalous
scaling is seen to be independent of dimension for $d \ge 3$,
although the actual existence of the steady state may very well
depend on the dimension as observed in \cite{arponen}. This will
be further studied in an undergoing investigation of the steady
state existence problem. The last section before the conclusion
attempts to shed light on the role of the parameter $a$ as it is
varied between $-1$ and $1$. The actual results are collected and
discussed in the conclusion. We also give some computational
details in the appendices.

\section{Preliminaries and the equation\label{setup}}
All vector quantities in the equation (\ref{equation}) being
divergence free results in an expression for the pressure after
taking the divergence,
\eq{P = (1-a) \left( - \Delta\right)^{-1} \partial_i v_j
\partial_j u_i.}
We may then write the equation compactly as 
\eq{\dot u_i - \nu \Delta u_i  + \mathcal D_{ijk} \left( u_j v_k
\right) = f_i,}
with a differential operator
\eq{\mathcal D_{ijk} = \delta_{ij} \partial_k - a \delta_{ik}
\partial_j + (a-1)\partial_i \partial_j \partial_k \Delta^{-1},}
where $\Delta^{-1}$ is the inverse laplacian. The equal time pair
correlation is defined as
\eq{G_{ij}(t,\mar) = \langle u_i (t,\maxx+\mar) u_j (t,\maxx)
\rangle,}
where the angular brackets denote an ensemble average with respect
to the forcing and the velocity field. The equation for the pair
correlation function is then
\eq{\partial_t G_{ij} -2 \nu' \Delta G_{ij} - \mathcal D_{i\mu \nu}
\mathcal D_{i\rho \sigma} \left( D_{\nu \sigma} G_{\mu \rho}
\right) = C_{ij},\label{paircorr}}
where the velocity and forcing pair correlation tensors $D_{ij}$
and $C_{ij}$ will be defined below. The above equation should
however be understood in a rather symbolic sense, as the defining
equation for the field $u$ is in fact a stochastic partial
differential equation. The equation is more carefully derived in
appendix \ref{APXeom} in Fourier variables using the rules of
stochastic calculus. In the present form it is also very difficult
to study because of the nonlocal terms for $a \neq 1$ and the
tensorial structure. We will therefore now briefly explain the structure of the calculations in a rather superficial but hopefully transparent way (see appendix \ref{APXeom} for details). Assuming that we have reached a steady state, i.e. $\partial_t G =0$, we rewrite eq. (\ref{paircorr}) symbolically as
\eq{-2 \nu' \Delta G + \mathcal M  G = C,}
with the effective diffusivity $\nu' = \nu-\frac{1}{2} D m_v^{-\xi}$ and $\mathcal M$ is some complicated integro-differential operator. Taking the Fourier transform of the above equation would still leave us with an integral equation due to the inherent nonlocality from the pressure term. We deal with this now by taking also the Mellin transform (after dividing by $p^2$), which yields
\eq{2 \nu' \bar g (z) + \inte \ud z' \mathcal M_{z,z'} \bar g (z-z') = \bar c (z-2)}
with a rather complicated expression for $\mathcal M_{z,z'}$, see eq. (\ref{A12}). The advantage of the above form is that various powers of $m_v$ arise as poles in $\mathcal M_{z,z'} \propto m_v^{z'-\xi}$, the leading ones residing at $z'=0$ and $z'=\xi$. Other poles produce positive powers of $m_v$ and can therefore be safely neglected. The residue at $z'=0$ cancels with the term in the effective diffusivity, leaving us with only the bare diffusivity $\nu$. The remaining equation can then be written in the limit of vanishing $m_v$ and $\nu$ as
\eq{-\mathcal R \left( \mathcal M_{z,z'} | z'=\xi \right) \bar g (z-\xi) = \bar c (z-2),}
where $\mathcal R$ denotes the residue (the minus sign arises from the clockwise contour). The equation is then simply solved by dividing by the residue term and using the Mellin transform inversion formula
\eq{G(\mar) = \inte \ud z |\mar|^z \mathcal A_z \bar g (z),}
where $\mathcal A_z$ is a simple $z$ dependent function arising from the fact that we performed the Mellin transform on the Fourier transform of the equation.
\subsection{Kraichnan model}
We define the Kraichnan model as in \cite{bernard} with the
velocity correlation function
\eq{ \big\langle  v_i(t,\mar)v_j(0,0)\big\rangle \!\!\!\! &&=
\delta(t) \inte \ud^d \maq e^{i \maq \cdot \mar}
\widehat{D}_{m_v}(q) P_{ij}(\maq)
\nonumber\\
 &&=: \delta(t) D_{ij}(\mar;m_v)\label{KraichCorr}}
where we have defined the incompressibility tensor
$P_{ij}(\maq)=\delta_{ij}-\widehat \maq_i \widehat \maq_j$ and
denoted $\ud^d \maq := \frac{d^d q}{(2 \pi)^d}$. Defining
\eq{ \widehat{D}_{m_v}(q) = \frac{\xi
D_0}{\left(\maq^2+m_v^2\right)^{d/2+\xi/2}} \label{massreg}}
and applying the Mellin transform (See e.g. \cite{mellin} and the
appendix A of \cite{paolo.antti}) we have
\eq{\widehat{D}_{m_v}^{z'}(\maq) := \int\limits_0^\infty
\frac{dw}{w} w^{z'+d} \widehat{D}_{m_v} (w  q) = \bar d_{m_v} (z')
q^{-z'-d}, }
where
\eq{\bar d_{m_v} (z') = \frac{\xi}{2} D_0 m_v^{z'-\xi}
\frac{\Gamma\left(d/2+z'/2 \right) \Gamma\left( \xi/2-z'/2
\right)}{\Gamma\left( d/2+\xi/2 \right)}, \label{dmv}}
and $z'$ is constrained inside the strip of analyticity $-d<
Re(z')<\xi$. The parameter $\xi$ takes values between zero and two
and measures the spatial "roughness" of the velocity statistics.
We observe that the scaling behavior of the correlation function
is completely encoded in the pole structure of Mellin transform,
with e.g. the pole at $z'=\xi$ corresponding to the leading
scaling behavior of the velocity structure function.
\subsection{Decomposition in basis tensor functions}
Being a rank two tensor field, the pair correlation function may
be decomposed in hyperspherical basis tensor functions as in
\cite{arad,arad2}. Such a decomposition is also an important tool
in analyzing the data from numerical simulations, as witnessed
e.g. in \cite{experimental}. We shall be concerned only with the
axial anisotropy, and apply this decomposition on the Fourier
transform of the pair correlation function. This has the advantage
of making the incompressibility condition very easy to solve,
among other things. We consider only the case of even parity and
symmetry in indices, which leaves us with a basis of four tensors:
\eq{  \ \left\{ \begin{array}{ll}
 B_{ij}^1 (\hat \map) &= |\map|^{-l} \delta_{ij} \Phi^l (\map) \\
 B_{ij}^2 (\hat \map) &= |\map|^{2-l} \partial_i \partial_j \Phi^l
(\map)
 \\
 B_{ij}^3 (\hat \map) &= |\map|^{-l} (p_i \partial_j + p_j
\partial_i)
 \Phi^l (\map) \\
 B_{ij}^4 (\hat \map) &= |\map|^{-l-2} p_i p_j \Phi^l (\map)
\end{array} \right. }
with the actual decomposition
\eq{\widehat{G}_{ij} (\map) := \sum\limits_{b,l} B_{ij}^{b,l}
(\hat \map) \widehat G_l^b (\map) .\label{covariances}}
Here $\Phi^l (\map)$ is defined as $\Phi^l (\map) := |\map|^l Y^l
(\hat \map)$, where $Y^l$ is the hyperspherical harmonic function
(with the multi-index $m=0$). It satisfies the properties
\eq{\Delta \Phi^l (\map) &&= 0 \nonumber \\ \map \cdot \nabla
\Phi^l (\map) &&= l \Phi^l (\map).\label{phiproperties}}
The same decomposition will naturally be applied to the forcing
correlation function as well.
\subsection{The forcing correlation function}
We require the forcing correlation function to decay faster than a
power law for large momenta and to behave as $C_{ij} (p) \propto
L^{d} (L p)^{2N}$ for small momenta with positive integer $N$. The
$N=0$ case corresponds to the usual large scale forcing with a
nonzero "charge" $q_0 = \inte d \mar C_L(\mar)$ and is responsible
for the canonical scaling behavior of the passive scalar at large
scales\cite{fouxon}, whereas any $N>0$ corresponds to a vanishing
charge\cite{fouxon}. Applying the Mellin transform to such a
tensor (decomposed as above) yields
\eq{\widehat{C}_{i j}^z (\map) &&= \int\limits_0^\infty
\frac{dw}{w} w^{d+z} \widehat{C}_{ij} (w \map)= |\map|^{-d-z}
\sum\limits_b B_{i j}^b (\hat \map) \bar c_{b}^N
(z),\label{mellins}}
with
\eq{\bar c_{b}^N (z) =   \frac{C_b^* L^{-z}}{z+d+2N}, \ \ \
\mathcal{R}e (z) > -d-2N, \label{cMf}}
and the strip of analyticity $-d -2N< \mathcal{R}e (z)$. The
details of the actual cutoff function are absorbed in the
constants $C_b^*$ and play no role in the leading scaling
behavior. All the interesting phenomena can be classified by using
only the cases $N=0$ and $N=1$. We will mostly be concerned with
the latter type of forcing which is also of the type considered in
\cite{celani,fouxon}. By inverting the Mellin transform we would
obtain an expression for the forcing correlation function
\eq{C_{ij}(t,\mar) &&= \inte \ud z  |\mar|^z \bar c_a^N (z)
\mathrm K^{ab}(z) B_{ij}^{b} (\hat \mar)}
where the matrix $\mathbf K$ is defined in appendix
\ref{Kappendix}. We note that $\bar c_a^N$ determines the large
scaling behavior of the above quantity as $r^{-d}$ or $r^{-2-d}$,
depending on the forcing, while the matrix $\mathrm K$ is
responsible for the small scale behavior $\propto r^{l}$, where
$l$ is the angular momentum variable.
\subsection{Mellin transformed equation and overview of calculations\label{steadystate}}
As mentioned earlier in this section, equation (\ref{paircorr}) is
much too unwieldy for actual computations. In appendix
\ref{APXeom} we perform a more careful derivation of the equation
in Fourier variables and by using the It\^o formula. The resulting
equation (\ref{fouriereq}) still has an inconvenient convolution
integral. By applying the Mellin transform, we obtain an equation
%
\eq{&& - D m_v^{-\xi} \bar g_b (z) - D_0 \widetilde \lambda \bar
g_b (z-\xi) + \inte \ud z' \bar d_{m_v}
(z') \mathrm T_{d+z',d+z-z'}^{bc} \bar g_c (z-z')  \nonumber \\
&&= \bar c_b (z-2).\label{mellineq}}
%
for the Mellin transformed coefficients $\bar g_b$ of the tensor
decomposition (\ref{covariances}) (defined explicitly in eq.
(\ref{mellinofcorr})). The matrix $\mathbf T$ is defined in eq.
(\ref{Lambda}) and involves rather difficult but manageable
integrals, and $\widetilde \lambda$ is defined in eq.
(\ref{lambda}). The integration contour with respect to $z'$ lies
inside the strip of analyticity $ \mathcal R e (z) < \mathcal R e
(z') < 0$, determined from eq. (\ref{strips1}). For small values
of $m_v$ the contour may (and must) be completed from the right.
The reason for performing the Mellin transform becomes evident
when one studies the pole structure of the functions $\bar d_{m_v}
(z')$ and $\mathbf T$: first two (positive) poles occur at $z'=0$
(from $\mathbf T$) and at $z'=\xi$ (from $\bar d_{m_v} (z')$) and
correspond to a term $\propto m_v^{-\xi}$ and a constant in $m_v$,
respectively. The former of these cancels out from the equation,
hence one is free to take the limit $m_v \to 0$. This leaves us
with a simple equation
\eq{&&- \widetilde \lambda \bar g_b (z-\xi) -  \mathrm
T^{bc}_{d+\xi,d+z-\xi} \bar g_c (z-\xi)= \frac{1}{D_0}\bar c_b
(z-2).}
From now on we absorb $D_0$ in the functions $\bar c_b$. In
appendix \ref{APXincomp} we have applied the incompressibility
condition to the correlation function $\widehat G_{ij} (\map)$ and
the equation, which has the effect of leaving us only with two
independent functions to be solved, $\bar g_1$ and $\bar g_2$.
Applying also a translation $z \to z+\xi$ in eq. (\ref{eq2}), we
have
\eq{-\left(\widetilde \lambda \mathbf{1} + \mathbf A+ \mathbf B
\cdot \mathbf X \right) \mathbf{\bar h} (z) = \mathbf{\bar f}
(z+\xi-2),\label{eq2}}
with the definitions
\eq{\mathbf{\bar h} &&= \left(\bar g_1 ,\bar g_2 \right)^{\mathrm
T} \nonumber \\ \mathbf{\bar f} &&= \left(\bar c_1 ,\bar c_2
\right)^{\mathrm T},}
and
\eq{ \mathbf T_{d+\xi,d+z} =
\begin{pmatrix}
 \mathbf A & \mathbf B \\ \mathbf C & \mathbf D
\end{pmatrix} \ \ \ , \ \ \ \mathbf X = \begin{pmatrix}
 0 & -(l-1) \\
 -1 & l(l-1)
\end{pmatrix}.}
All that remains now is to invert the matrix equation, although in
the isotropic sector and in two dimensions it reduces to a scalar
equation.

\section{The solution\label{solution}}
Inverting the Mellin and Fourier transforms enables us to write
the full solution as
\eq{G_{ij}  (\mar)= - \inte \ud z |\mar|^z \mathbf{\bar
h}^{\mathrm T} (z) \mathbf{\widehat P}^{\text T} \mathbf K \cdot
B_{ij} (\hat \mar),\label{correlation}}
where we now have a projected version of the matrix $\mathbf K$
due to the incompressibility condition (see appendix
\ref{Kappendix}), and
\eq{\mathbf{\bar h} (z) =-\left(\widetilde \lambda \mathbf{1} +
\mathbf A+ \mathbf B \cdot \mathbf X \right)^{-1} \mathbf{\bar f}
(z+\xi-2).}
The strip of analyticity is now
\eq{2-d-\xi -2N < \mathcal R e (z) < 0,}
where $N=0$ for the traditional nonzero charge forcing and $N=1$
for the zero charge forcing. We should note that there may in fact
be poles inside the strip of analyticity due to the solution $\bar
{\mathbf{h}}$, which is just a reflection of one's choice of
boundary conditions.

\subsection{Isotropic sector}
In the isotropic case when $l = 0$, we have $B^1_{ij} =
\delta_{ij}$, $B^4_{ij} = \widehat{\mathbf r}_i \widehat{\mathbf
r}_j$ and the other $B$'s are zero. The equation of motion
(\ref{eq2}) is now a scalar equation, hence we only need the
$(1,1)$ -component of the matrix,
%
\begin{widetext}
\eq{&&\left(\widetilde \lambda \mathbf{1} + \mathbf A+ \mathbf B
\cdot \mathbf X \right)_{11} = \frac{d}{2} (a-1) (a \xi -1 - a -
d) \Gamma \left( 1+\xi/2 \right) \Gamma \left( 1+d/2 \right)\nonumber \\
&& -2 p_a(z) \frac{\Gamma \left( -z/2 \right) \Gamma \left(
1-\xi/2 \right) \Gamma \left( \frac{d+z+\xi}{2} \right)
  \Gamma \left( \frac{4+d-\xi}{2} \right)}{\Gamma \left(
  \frac{2+d+z}{2} \right)
   \Gamma \left( \frac{4-z-\xi}{2} \right) } \doteq 1/\gamma_a(z),\label{1overgamma}}
where the equality applies up to a constant term that will be
absorbed in the forcing, and we have defined the polynomial
\eq{&&p_a(z) = (a-1)^2 (1+d)\xi z \nonumber \\ && +(z+\xi-2)
\left((d-1) z (d+z) +a (a (d-1) d+2 z) \ \xi+a^2 (d-1)
\xi^2\right)}
\end{widetext}
This is the same expression (only in a slightly different form) as
in \cite{paolo}. The expression (\ref{correlation}) for the
inhomogeneous part of the correlation function becomes
\eq{ G_{ij}(\mar)  =\!\!\!\!\!\!\! &&\inte \ud z |\mar|^z \gamma_a
(z) c_1 (z+\xi-2) \mathcal P_{ij} (z) \frac{\Gamma(-z/2)}{\Gamma
\left( \frac{2+d+z}{2} \right)} ,\label{isotropiccorrelation}}
where we have introduced the incompressibility tensor
\eq{\mathcal P_{ij} (z) = \left[ (z+d-1) \delta_{ij}-z
\hat{\mathbf r}_i \hat{\mathbf r}_j
\right]\label{incompressibilitytensor}}
and irrelevant constant terms were absorbed in the forcing
$c_1$.Henceforth such an assumption will always be implied unless
stated otherwise.
\subsection{Anisotropic sectors}
Now the task is to find the poles of the inverse matrix of $\left(
\widetilde \lambda\mathbf{1}+\mathbf A+ \mathbf B \cdot \mathbf X
\right)$, that are completely determined by the zeros of its
determinant. Denoting
\eq{&&\mathbf M:=\mathbf A+\mathbf B \cdot \mathbf X =
\frac{\lambda_{l+d+z,d+\xi}}{d+\xi}\begin{pmatrix}
 \tau_{11} -\tau_{41}  & \tau_{21} - (l-1) \tau_{31} +
 (l-1) l \tau_{41} \\
 \tau_{12} -\tau_{42}  & \tau_{22} - (l-1) \tau_{32} +
 (l-1) l \tau_{42}
\end{pmatrix} ,}
where $\tau$ and $\lambda$ are defined in appendix
\ref{tauappendix}, we may write
\eq{\det \left( \widetilde \lambda\mathbf{1}+\mathbf M \right) =
{\widetilde \lambda}^2 + \widetilde \lambda \ \text{tr} \mathbf M
+ \det \mathbf M.\label{determinant}}
We refrain from explicitly writing down the determinant, since the
full expression is rather cumbersome and not very illuminating. It
may however be easily reproduced by using the components
$\tau_{ij}$ given in appendix \ref{tauappendix}.
\subsection{Two dimensions}
The two dimensional case deserves some special attention. From the
incompressibility requirement in eq. (\ref{incompcond}) and by
direct computation using the two dimensional spherical harmonics
$\propto e^{\i \theta}$, one can see that the correlation function
satisfies the propotionality
\eq{\widehat G_{ij} (\map) \propto \left(\bar g^1 - l(l-1) \bar
g^2 \right) P_{ij} (\map).}
Therefore in two dimensions the equation is a scalar one also in
the anisotropic sectors. A formula for the solution then becomes
\eq{G_{ij}  (\mar)= -  \inte \ud z |\mar|^z \frac{\bar c^1
-l(l-1)\bar c^2}{\text{F}_{11}-l(l-1) \text{F}_{21}}
(\mathbf{\widehat P}^{\text T} \mathbf K)^{1b} B_{ij}^b (\hat
\mar),}
where $\mathbf{F} = \widetilde \lambda \mathbf{1} + \mathbf A+
\mathbf B \cdot \mathbf X $.
\subsection{Example: Passive Scalar \label{passivescalar}}
As one of the main themes of the present work is to consider the
effect of a forcing localized around some finite wavenumber $m_f
\propto 1/L$ instead of zero, it is useful to review the case in
\cite{celani} by the present method  (see e.g.
\cite{gawedzkikupiainen,paolo.antti,bernard} for more on the
passive scalar problem), even more so as the magnetohydrodynamic
case in two dimensions bears close resemblance to the passive
scalar (indeed the two dimensional case can be completely
described as a passive scalar problem with the stream function
taking place of the scalar). Using the methods above, we arrive at
an expression similar to (\ref{correlation}),
\eq{ &&G (\mathbf{\mar})= \sum\limits_l Y_l ( \mathbf{\mar}) \inte
\ud z \left| \mar \right|^z \frac{c^N (z+\xi-2)}{\psi_l (z)}
\frac{\Gamma \left( \frac{l-z-\xi+2}{2}\right)}{\Gamma \left(
\frac{l+z+d+\xi-2}{2}\right)} ,}
where we have again written the generic constant $C'$ in which we
will absorb finite constants. In the above equation, $N$ equals
zero or one corresponding to the nonzero and zero charge forcings
and
\eq{&&\psi_l(z) =   (d-1)(l-z)(l+z+d+\xi-2) + \xi l (l-1) .
\label{psi}}
The strip of analyticity is now $-d-\xi < \mathcal R e (z) < 0$.
Consider now the isotropic sector $l=0$ with the nonzero charge
forcing, i.e. $N=0$. We then have (neglecting the zero modes)
\eq{G_{l=0} (\mathbf{x}) = C' L^{2-\xi} \inte \ud z \left|
\mathbf{\mar}/L \right|^z  \frac{\Gamma \left(
\frac{2-z-\xi}{2}\right)}{z \left( z+d+\xi-2\right)}.}
For $r \ll L$ the integration contour must be completed from the
right, thus capturing the poles $z=0, z= 2-\xi, \ldots$. The small
scale leading order behavior is therefore
\eq{G_{l=0} = C' \frac{\Gamma \left( 1-\xi/2 \right)}{d+\xi-2}
L^{2-\xi} - C' \frac{1}{d (2-\xi)} r^{2-\xi} + \ldots}
where the dots refer to higher order powers of $r$. The large
scales $r \gg L$ require a left hand contour, resulting in another
scaling regime,
\eq{G_{l=0} = C' \frac{\Gamma (d/2)}{d+\xi-2} L^{d} r^{2-d-\xi} +
\ldots}
We note that the above solution is constant at $r=0$ and zero at
$r=\infty$, thus satisfying the boundary conditions. We conclude
that the solution is completely nonanomalous, i.e. respecting the
canonical scaling.
\\ \\
Consider now instead the zero charge forcing with $N=1$ that is
localized around $p=1/L$ instead of $p=0$. The large scale pole
due to the forcing at $z=-d-\xi + 2$ cancels out and we are left
with
\eq{G_{l=0} = C' \inte \ud z \left| \mar/L \right|^z \frac{1}{z}
\Gamma \left(\frac{2-z-\xi}{2} \right).}
There is now no large scale scaling behavior (the decay is faster
than a power law). By looking at the $l=2$ sector,
\eq{G_{l=2}  = C'' L^{2-\xi} \inte \ud z \frac{\left| \mar/L
\right|^z\Gamma \left( \frac{4-z-\xi}{2}\right)}{ \psi_2 (z)
(z+\xi+d) \Gamma \left( z+d+\xi\right)},}
(with a different generic constant $C''$), we see that the
relevant scaling behaviors are obtained from a solution of the
equation
\eq{\psi_2 (z) = d^2 (-2 + z) - z (-2 + z + \xi) + d (-2 + z) (-1
+ z + \xi) = 0,}
giving the large scale behaviour of the $l=2$ sector with the
exponent
\eq{z_- = \frac{1}{2} \left(2 - d - \xi -
   \sqrt{(d - 2 + \xi)^2 + \frac{8 d (d + \xi - 1)}{d - 1}}\right).}
Therefore we conclude that the large scale behavior is dominated
by the anisotropic modes. Note that the anisotropic modes are also
anomalous in that they are not obtainable by dimensional analysis.
\section{Magnetohydrodynamic turbulence}
Setting $a=1$ in eq. (\ref{equation}) yields the equations of
magnetohydrodynamic turbulence (see e.g. \cite{vergassola,arponen}
and references therein). This is a special case in that the
problem is completely local due to the vanishing of the pressure
term. In practical terms, the quantity $\widetilde \lambda$ is
zero, hence we only need to consider the zeros of the determinant
of $\mathbf M$ in eq. (\ref{determinant}).
\subsection{Isotropic sector}
The isotropic part of the correlation function becomes
\eq{G_{ij}(\mar)\! =\! C' \inte \ud z |\mar|^z \frac{c_L^N
(z+\xi-2)}{p_0(z)} \mathcal P_{ij}(z) \frac{\Gamma \left(
\frac{2-z-\xi}{2} \right)
 }{\Gamma \left( \frac{d+z+\xi}{2} \right)} ,}
where
\eq{&&p_0 (z) = (d-1) z (d+z)+((d-1) d+2 z) \xi +(d-1) \xi^2}
with another generic constant $C'$. We find the usual poles at
\eq{z_n &&= 2-\xi + 2 n, \nonumber \\
    z_{\pm} &&= \frac{1}{2}\left( -d -\frac{2\xi}{d-1} \right)
    \pm
 \frac{\sqrt{d}}{2}  \sqrt{
  d - \frac{4 (d - 2) \xi}{(d - 1)} - \frac{4 (d - 2) \xi^2}
  {(d - 1)^2}} ,}
where $n$ is a nonnegative integer. For the nonzero charge type
forcing we have $c_L^0(z+\xi-2) \propto 1/(z+d+\xi-2)$, which
presents another pole. On the other hand, for the zero charge
forcing we have $c_L^1(z+\xi-2) \propto 1/(z+d+\xi)$, which
cancels with a zero of the gamma function. It turns out that this
sort of cancelation occurs for each model, rendering the large
scale behavior anomalous. We will postpone the arbitrary
dimensional case until the end of the present sector and instead
consider first the three and two dimensional cases.
\subsection{Anisotropic sectors}
Note that since $\det \mathbf M \propto \lambda_{l+d+z,d+\xi}^2$,
the inverse of $\mathbf M$ is only proportional to
$\lambda_{l+d+z,d+\xi}^{-1}$, so the correct form to look at is
actually $\det M /\lambda_{l+d+z,d+\xi}$. Dropping $z$
-independent terms we have
\eq{\frac{\det \mathbf M}{\lambda_{l+d+z,d+\xi}} = C \frac{\Gamma
\left( \frac{l-z-2}{2} \right)\Gamma \left(\frac{l+z+d+\xi-2}{2}
\right)}{\Gamma \left( \frac{l+z+d+2}{2} \right)\Gamma
\left(\frac{l-z-\xi+2}{2} \right)} \Psi_l (z),}
where $\Psi_l (z)$ is a fourth order polynomial in $z$ and $C$ is
a $z$ -independent constant. Due to its rather lengthy expression,
we shall consider the whole problem in three and two dimensions
only. We note immediately that there's also an infinite number of
solutions due to one of the gamma functions, namely at
\eq{z= l+2-\xi + 2 n}
for nonnegative integers $n$ and even $l$. The other gamma
function cancels with the terms from (\ref{PTK}).
\subsection{d=3}
\begin{figure*}

\begin{center}
\subfigure[\ d=3]{
\includegraphics[scale=0.48]{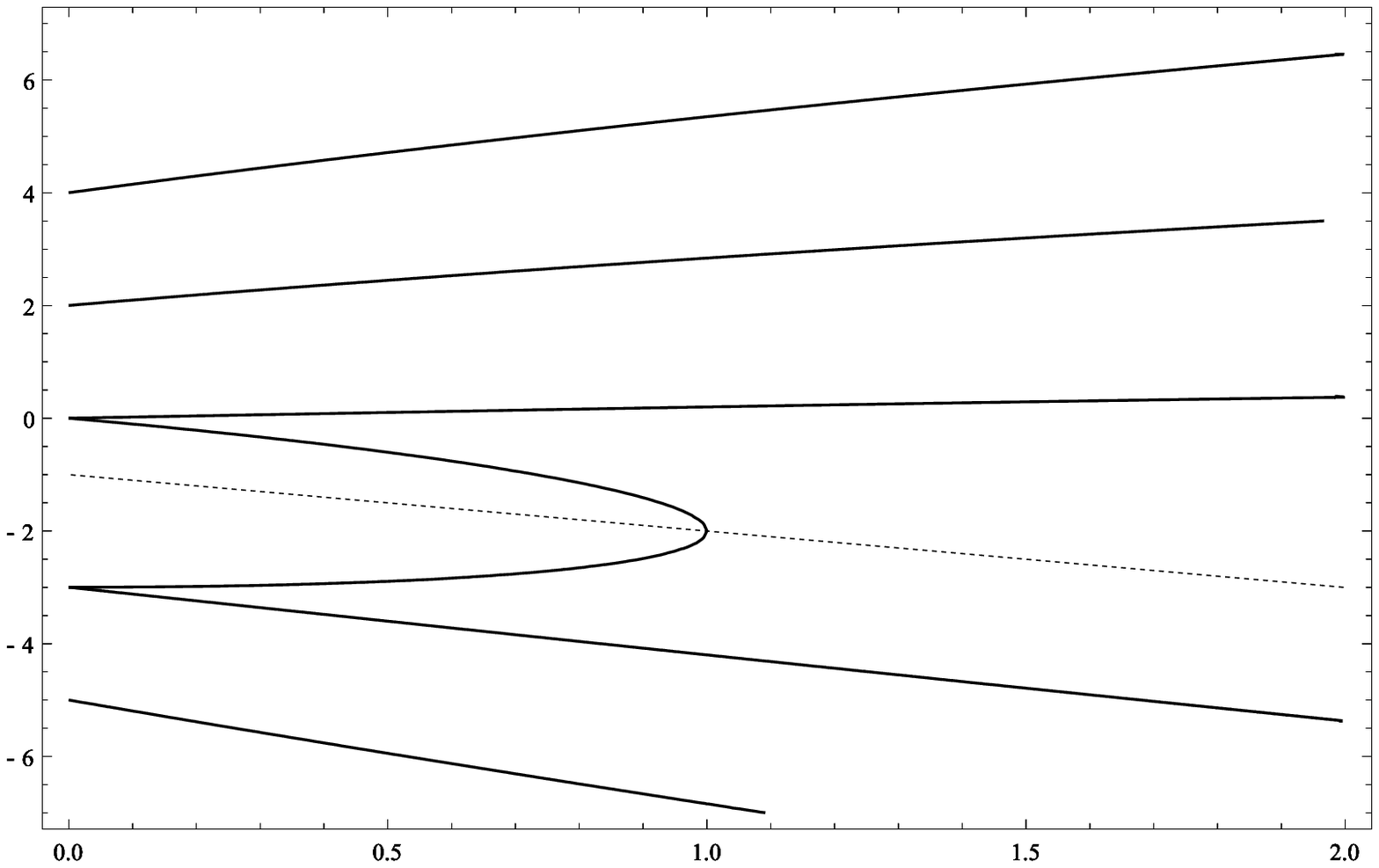}
\put(-30,30){\footnotesize{$l=2$}}
\put(-30,62){\footnotesize{$l=2$}}
\put(-30,94){\footnotesize{$l=4$}}
\put(-30,120){\footnotesize{$l=6$}}
\put(-200,9){\footnotesize{$l=4$}} } \subfigure[\ d=2]{
\includegraphics[scale=0.48]{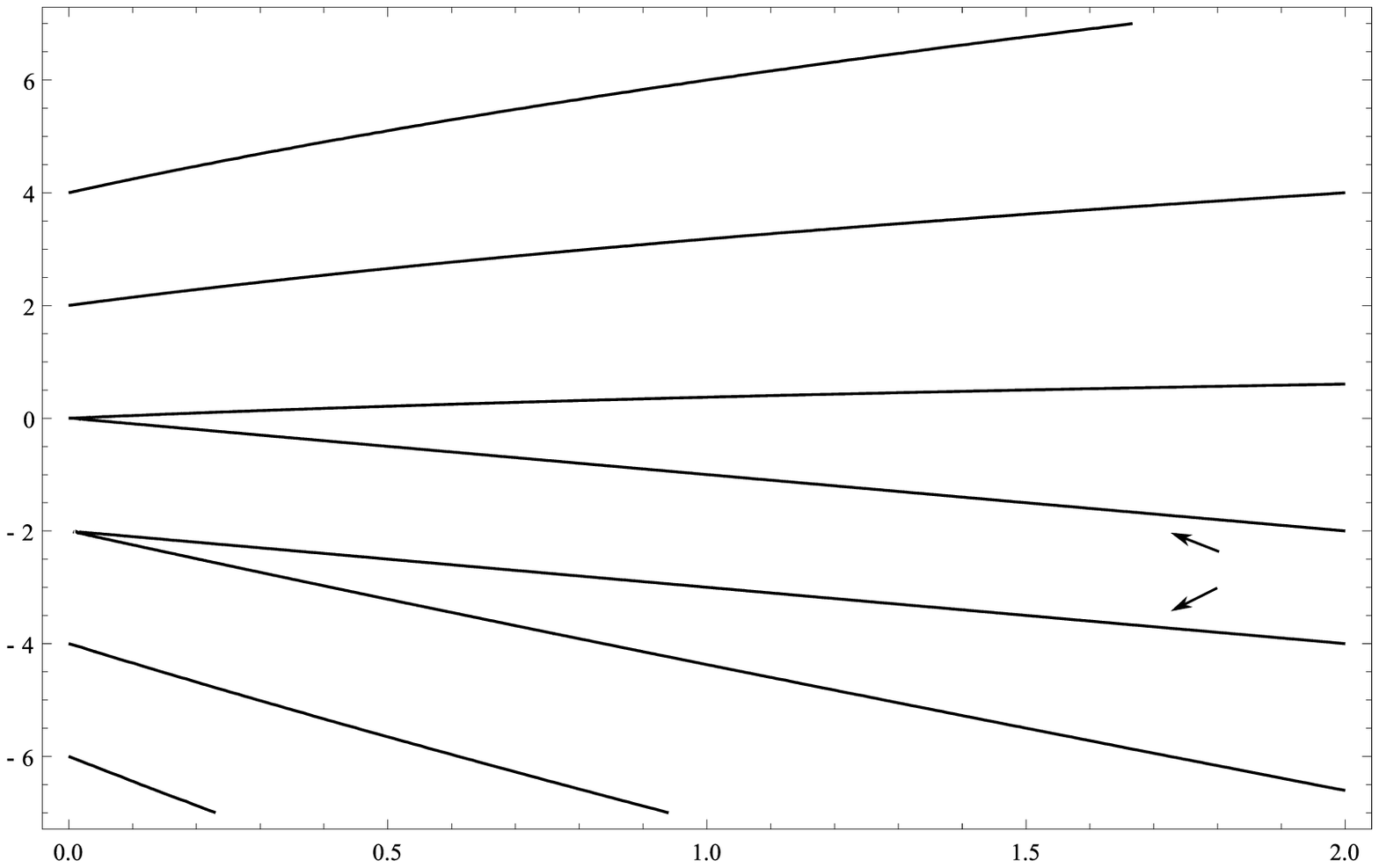}
\put(-30,20){\footnotesize{$l=2$}}
\put(-23,45){\footnotesize{$l=0$}}
\put(-30,68){\footnotesize{$l=2$}}
\put(-30,98){\footnotesize{$l=4$}}
\put(-40,127){\footnotesize{$l=6$}}
\put(-118,10){\footnotesize{$l=4$}}} \caption{\label{MHDL0toL6}The
MHD scaling exponents of the isotropic, $l=2$, $l=4$ and $l=6$. In
(a) The isotropic poles $z_+ \ge z-$ are adjoined at $\xi=1$. The
dashed line in (a) corresponds to the forcing with nonzero charge
with a pole at $-1-\xi$, whereas for the zero charge forcing there
are no poles. In (b) the zero modes are never adjoined.}
\end{center}
\end{figure*}

We have the four solutions to $\Psi_l (z) = 0$ of which the
following two are dominant in the small and large scales,
\eq{z^\pm_l = -\frac{3+\xi}{2} \pm \frac{1}{2} \sqrt{A \mp 2
(2-\xi) \sqrt{B}},}
where \eq{A &= (2+\xi)\left( 2 l (l+1)-6-\xi \right)+17, \nonumber
\\ B &= (2+\xi)\left( 2l (l+1) + \xi \right)+1,}
which match exactly to the results obtained in
\cite{lanotte2,arad}, after some convenient simplifications. The
isotropic zero modes are
\eq{z_\pm = \frac{1}{2}\left(  \-3-\xi \pm \sqrt{3(1-\xi)(3+\xi)}
\right).}
We have plotted the leading poles in Fig.\ (\ref{MHDL0toL6}) from
$l=0$ to $l=6$ together with the pole due to the nonzero charge.
We note that the isotropic exponents become complex valued for
$\xi>1$, implying an oscillating behavior and therefore a positive
Lyapunov exponent for the time evolution
\cite{vergassola,arponen}. The above steady state assumption
therefore applies for $0\le \xi \le1$ only in the isotropic
sector. The fact that the anisotropic exponents are continuous
curves for all $0 \le \xi \le 2$ seems to imply that the steady
state exists for all $\xi$ in the anisotropic sectors. Indeed, in
\cite{arad} this was shown to be the case by preforming a more
careful eigenvalue analysis.
\subsubsection{Nonzero forcing charge}
In the isotropic sector for the forcing with nonzero charge $N=0$
we have
\eq{G_{ij}(\mar)|_{l=0} &&= - C' L^{2-\xi} \inte \ud z
 \frac{|\mar/L|^z \mathcal P_{ij}(z) \Gamma \left(
\frac{2-z-\xi}{2} \right)}{(z-z_-)(z-z_+)(z+1+\xi)\Gamma \left(
\frac{3+z+\xi}{2} \right)}}
with the contour bound $-1-\xi < \mathcal R e (z)<0$ and $z_- <
-1-\xi < z_+ < 0$. $C'$ again denotes some generic finite (and
positive) constant. The pole $z_+$ divides the strip of
analyticity in two parts, which correspond to different boundary
conditions. Small scale behavior corresponds to picking up the
poles to the right of the contour and large scale behavior
corresponds to left hand side poles. We note that both the zero
modes $z_\pm$ are negative, except that $z_+ = 0$ at $\xi=0$.
Therefore $z_+$ cannot be a large scale exponent, as the solution
has to decay at infinity. The real strip of analyticity is then in
fact $-1-\xi < \mathcal R e (z) < z_+$, thus resulting in the
small scale behavior
\eq{G_{ij}^< = C_1 r^{z_+} \mathcal P_{ij} (z_+)}
and the large scale behavior
\eq{G_{ij}^> = C_2 r^{-1-\xi} \mathcal P_{ij} (-1-\xi)}
We note that the large scale behavior is determined by the forcing
and therefore respects canonical scaling.
\subsubsection{Zero charge forcing}
Because of the pole cancelation we now have a similar expression,
\eq{G_{ij}(\mar)|_{l=0} &&=  C' L^{2-\xi} \inte \ud z
\frac{|\mar/L|^z \mathcal P_{ij}(z)\Gamma \left( \frac{2-z-\xi}{2}
\right)}{(z-z_-)(z-z_+)\Gamma \left( \frac{5+z+\xi}{2} \right)}}
with the strip of analyticity is now $-3-\xi < \mathcal R e
(z)<0$. The contour bound now encloses both the zero modes (see
again Fig.\ (\ref{MHDL0toL6})). In addition to the above
considerations with a forcing of nonzero charge, we conclude that
$z_-$ cannot be present at small scales due to regularity
conditions at $\xi = 0$, so the real strip of analyticity is in
fact $z_- < \mathcal R e (z)<z_+$. This gives rise to the small
scale behavior
\eq{G_{ij}^< = C_1 r^{z_+} \mathcal P_{ij} (z_+)}
for the small scales and
\eq{G_{ij}^> = C_2 r^{z_-}\mathcal P_{ij} (z_-)}
for the large scales. The large scales are therefore dominated by
the smaller zero mode $z_-$ instead of the exponent $-1-\xi$ as
with the nonzero charge forcing and is therefore anomalous.
However, unlike in the passive scalar case, the anisotropic
exponents are subdominant at both small and large scales (see
Fig.\ (\ref{MHDL0toL6})) and we therefore conclude that there is
isotropization at both scales.
\subsection{d=2}
The (dominant) zero modes in two dimensions are
\eq{z_l^+ &= -4-\xi + \sqrt{4 l^2 (1+\xi) + \xi^2} \nonumber \\
    z_l^- &= -3\xi - \sqrt{4l^2(1+\xi) + \xi^2}}
of which we separately mention the isotropic zero modes,
\eq{z_+ &= -\xi \nonumber \\
    z_- &= -2-\xi.}
The expression for the inhomogeneous part of the correlation
function is
\eq{G_{ij}(\mar)|_{l=0} &&= C' L^{2-\xi} \inte \ud z |\mar/L|^z
c_L (z+\xi-2) \mathcal P_{ij}(z)
 \frac{\Gamma \left(
\frac{-z-\xi}{2} \right)
 }{\Gamma \left( \frac{4+z+\xi}{2} \right)} }
with contour bound is $-2-\xi< z <0$ together with the bound from
the forcing.
\subsubsection{Nonzero charge forcing}
Because $c_L \propto 1/(z+\xi)$, the expression for the isotropic
sector of the correlation function simplifies to
\eq{G_{ij}(\mar)|_{l=0} &&= C' L^{2-\xi}  \inte \ud z |\mar/L|^z
\mathcal P_{ij}(z)
 \frac{\Gamma \left(
\frac{-z-\xi}{2} \right)
 }{(z+\xi)\Gamma \left( \frac{4+z+\xi}{2} \right)}  }
where the bound is now $-\xi < \mathcal R e (z) < 0$. Note the
appearance of a double pole at $z=-\xi$ giving rise to logarithmic
behavior. There are now no poles inside the contour bound, so
finding the asymptotics is easy. We observe that there are no
small scale poles and therefore the correlation function decays
faster than any power at small scales, whereas at large scales we
have
\eq{G_{ij}^> &&= C' \log (r/L) L^{2-\xi} |r/L|^{-\xi} \mathcal
P_{ij} (-\xi) + C'L^{2-\xi} |r/L|^{-\xi} \mathcal P'_{ij} (-\xi),}
where  $\mathcal P'_{ij} (-\xi) = \delta_{ij}- \hat{\mathbf r}_i
\hat{\mathbf r}_j$ to ensure incompressibility and other next to
leading order nonlogarithmic terms were discarded. By looking at
Fig. (\ref{MHDL0toL6}) we see that there is a hierarchy of small
scale exponents in the anisotropic sectors. We therefore make the
conclusion that in two dimensions the anisotropic effects in the
MHD model are dominant at small scales for a forcing of
nonvanishing charge, conversely to the passive scalar case. Note
that setting $\xi=0$ in the above equation reproduces correctly
the usual logarithmic behavior of the diffusion equation steady
state with an infrared finite large scale forcing.
\subsubsection{Zero charge forcing}
We now have $c_L \propto 1/(z+\xi+2)$ and the isotropic
correlation function becomes
\eq{G_{ij}(\mar)|_{l=0} &&= - C' L^{2-\xi} \inte \ud z |\mar/L|^z
\mathcal P_{ij}(z)
 \frac{\Gamma \left(-1-
\frac{z+\xi}{2} \right)
 }{\Gamma \left( \frac{4+z+\xi}{2} \right)}  }
with the usual strip $-2-\xi < \mathcal R e (z) < 0$. There are no
double poles and the leading simple poles are just at $z=- \xi$
and $z=-2-\xi$, so the asymptotic behaviours at small and large
scales are simply
\eq{G_{ij}^< &&= C'  |r/L|^{-\xi} \mathcal P_{ij} (-\xi)\nonumber \\
    G_{ij}^> &&= C'  |r/L|^{-2-\xi} \mathcal P_{ij} (-2-\xi).}
As in the three dimensional case, all the anisotropic exponents
are now subleading at both small and large scales (see Fig.\
(\ref{MHDL0toL6})), so we conclude that there is again
isotropization at both regimes. Note also that the large scale
behavior is due to the forcing and therefore nonanomalous.
\subsubsection{Any dimension, zero charge forcing}
For the sake of completeness, we write explicitly the solutions in
any dimension $d>2$ in the isotropic sector for the zero charge
forcing:
\eq{G_{ij}^< &&= \frac{C'}{2} L^{2-\xi} |r/L|^{z_+} \frac{\mathcal
P_{ij} (z_+)}{z_+-z_-} \frac{\Gamma \left( \frac{2-z_+
-\xi}{2}\right)}{\Gamma \left( \frac{2+d+z_+ + \xi}{2}\right)}
\nonumber \\
&&- C' r^{2-\xi}  \frac{\mathcal P_{ij}
    (2-\xi)}{(2-\xi-z_-)(2-\xi-z_+)}
    \frac{1}
    {\Gamma \left( 2+d/2\right)}+ \mathcal O (r^{4-\xi}), \nonumber
\\
    G_{ij}^> &&= \frac{C'}{2} L^{2-\xi}
    |r/L|^{z_-} \frac{\mathcal P_{ij} (z_-)}
    {z_+-z_-} \ \frac{\Gamma \left( \frac{2-z_- -\xi}
    {2}\right)}{\Gamma \left( \frac{2+d+z_- + \xi}{2}\right)}}
where we have neglected the possible exponentially decaying terms.
The anisotropic sectors produce rather cumbersome expressions and
we will be satisfied with only the numerical results in the
figures. We observe that the large scale behavior is always
dominated by the negative zero mode exponent $z_-$ and is
therefore always anomalous (except in two dimensions). It is also
fairly easy to see that the anisotropic exponents are always
subdominant, so that there is isotropization at both small and
large scales.
\section{Linear Pressure Model}
Setting $a=0$ in eq. (\ref{equation}) produces the equation known
as the Linear Pressure Model (LPM) (see e.g.
\cite{paolo,adz,jurc1} and references therein; sometimes this
model is just called the passive vector model) By looking at
equation (\ref{fouriereq}), we see that when $\widehat G \propto
\delta^{(d)} (\map)$, the left hand side evaluates to $\propto a^2
|\map|^{2-d-\xi} P_{ij}(\map)$. Therefore for $a=0$ there is a
constant zero mode analogously to the passive scalar case. This is
true for the anisotropic sectors as well (\cite{adz,jurc1}). This
constant zero mode however vanishes for the structure function, so
in the present case we also consider the next to leading order
term.
The first thing to note in the isotropic sector is that when
$a=0$, $z= -d$ is a solution of the equation
\eq{&&\frac{1}{\gamma_0(z)} =   \frac{\xi}{8} (d+1)\Gamma\left( \xi/2 \right) \Gamma \left(1 + d/2 \right) \nonumber \\
&&+ p_0(z) \frac{\Gamma \left( -z/2 \right)  \Gamma \left(
\frac{d+z+\xi}{2} \right)  \Gamma \left( \frac{4+d-\xi}{2}
\right)}{\Gamma \left( \frac{2+d+z}{2} \right) \Gamma \left(
\frac{4-z-\xi}{2} \right) }=0,\label{gammazero}}
where
\eq{&&p_0 (z) = z (d^2 (z + \xi - 2) - z (z + \xi - 2) +
   d (z - 2) (z + \xi - 1) - \xi).}
However, as we see from the definition of the incompressibility
tensor in eq. (\ref{incompressibilitytensor}), for the trace (in
indices) we have
\eq{\mathcal P_{ii} (z) = (d-1)(d+z),}
which produces a canceling $z+d$ term in the numerator. A
physically more realistic quantity would however be a contraction
with $\widehat \maxx^i \widehat \maxx^j$ than the trace, since we
are more interested in the structure functions of the model.
Another exact solution is $z= 2 -\xi$. Other nonperturbative
solutions can only be obtained numerically.
\subsection{Any dimension}
\begin{figure*}
\begin{center}
\subfigure[\ d=3]{
\includegraphics[scale=0.48]{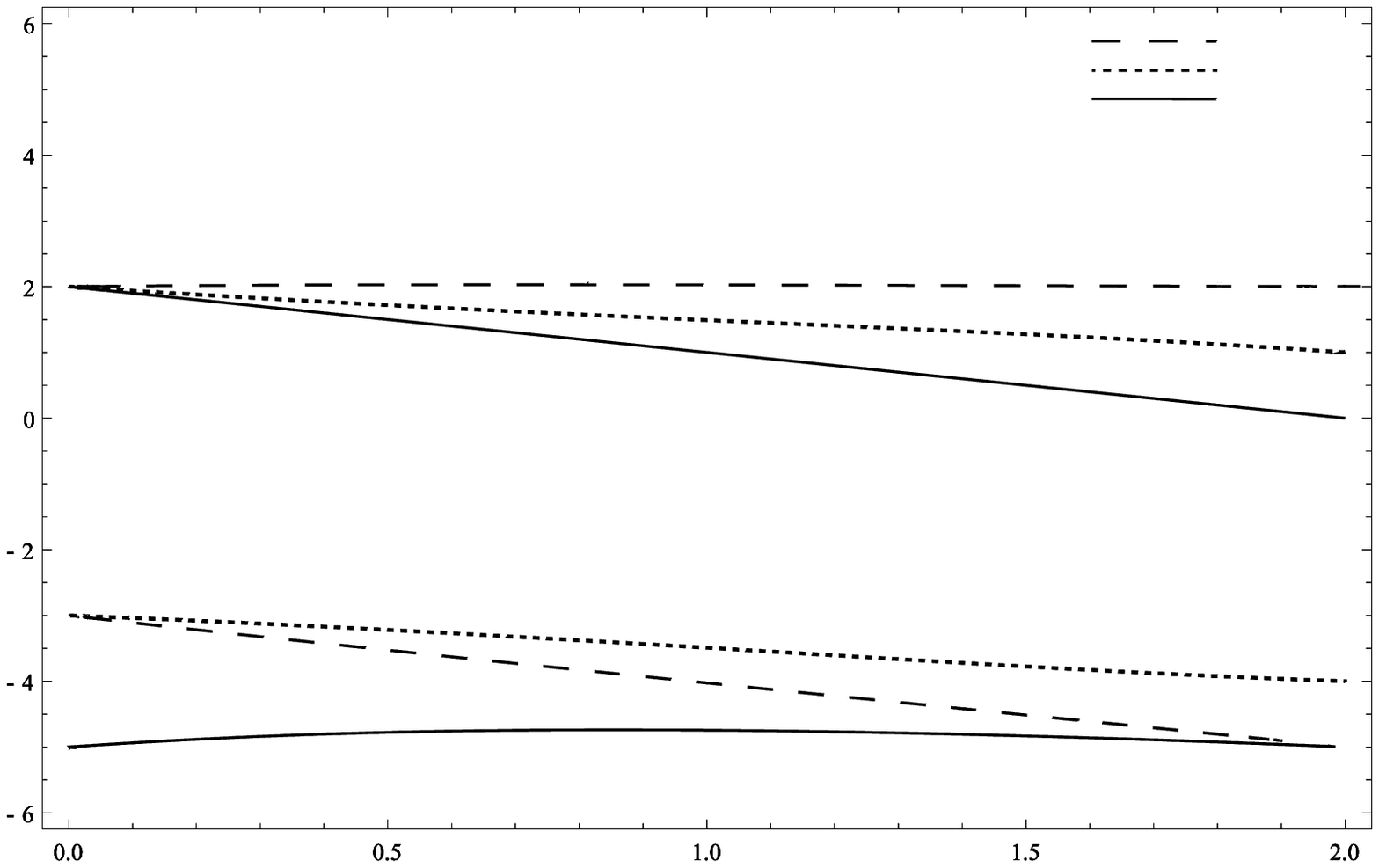}
\put(-22,132){\tiny{$l=4$}} \put(-22,127){\tiny{$l=2$}}
\put(-22,122){\tiny{$l=0$}}} \subfigure[\ d=2]{
\includegraphics[scale=0.48]{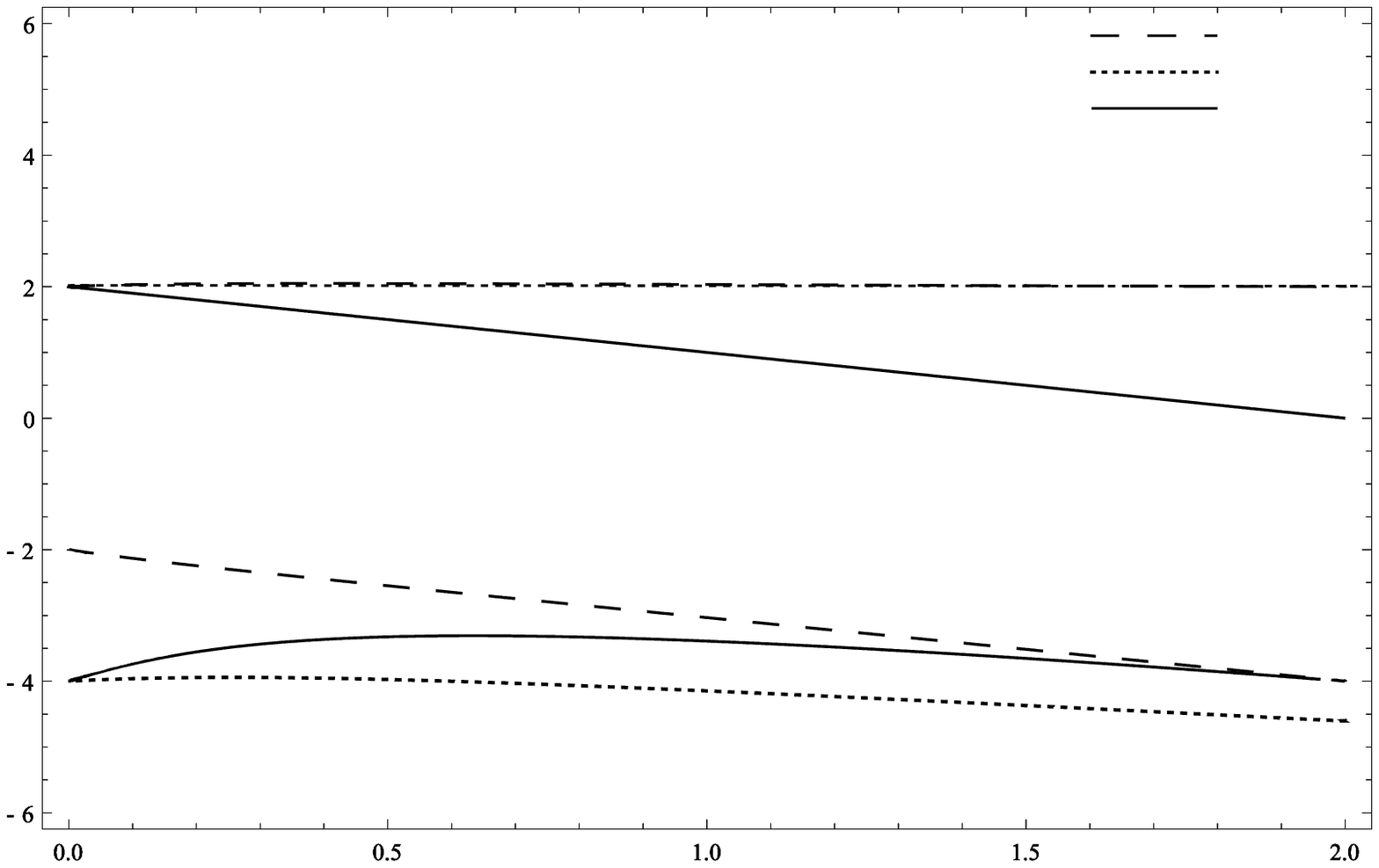}
\put(-22,132){\tiny{$l=4$}} \put(-22,127){\tiny{$l=2$}}
\put(-22,122){\tiny{$l=0$}}} \caption{The Linear Pressure Model
scaling exponents of the sectors $l=0$, $l=2$ and $l=4$ in three
(a) and two (b) dimensions. The $z=0$ and $z=-d$ are omitted for
the sake of clarity. We note that in two dimensions, there is a
$z=2$ exponent in the $l=2$ sector but the $l=4$ sector's exponent
goes slightly above $z=2$.\label{PVL0toL4}}
\end{center}
\end{figure*}
We have plotted some of the poles in Fig. (\ref{PVL0toL4}) in
three dimensions. Remembering the $z=-3$ solution, we see that the
anisotropic exponents are less dominant with increasing $l$ (a
behavior repeated for higher $l$ as well).
\subsubsection{Nonzero charge forcing}
The contour bound is now $2-d-\xi < \mathcal R e (z) < 0$, so
there is no controversy in the choice of which poles to include.
The small and large scale behaviors are similar to the passive
scalar, and for completeness, we give the results in any
dimension:
\eq{G_{ij}^< &&= A L^{2-\xi} \delta_{ij} - B r^{2-\xi}
\mathcal P_{ij} (2-\xi) \nonumber \\
    G_{ij}^> &&= A' L^{2-\xi} |r/L|^{2-d-\xi}\mathcal P_{ij} (2-d-\xi).}
The $A,B$ and $A'$ are somewhat complicated transcendental
functions of $d$ and $\xi$.
\subsubsection{Zero charge forcing}
Now the forcing contributes a pole $\propto 1/(z+\xi+d)$ and the
contour bound is $-d-\xi < \mathcal R e (z) < 0$. The quantity
$\gamma_0$ in eq. (\ref{gammazero}) has a zero there that cancels
with the pole from the forcing. Therefore we again conclude that
the forcing doesn't contribute in the scaling. The small scale
behavior is therefore same as above, but the large scale isotropic
sector of the correlation function behaves as
\eq{&&G_{ij}^> =  C' |L|^{2-\xi} \left( A' |r/L|^{-d} \mathcal
P_{ij} (-d) + B' |r/L|^{z_-} \mathcal P_{ij} ( z_-) \right)\!,
\label{largePV}}
where $A'$ and $B'$ are again some nonzero constants (depending of
$d$ and $\xi$), $z_-$ is the $l=2$ large scale mode (see Fig.
(\ref{PVL0toL4})) and we have the traceless tensor
\eq{\mathcal P_{ij} (-d) = d \hat \mar_i \hat \mar_j
-\delta_{ij}.}
By looking at Fig. (\ref{PVL0toL4}) we observe that the subleading
exponent $z_-$ is smaller than the anisotropic exponent $l=2$ in
three dimensions and $l=4$ at two dimensions (except when $\xi$ is
close to two, when the $l=2$ exponent is larger than the $l=4$
exponent). Therefore the \emph{trace} of the correlation function
is dominated by the anisotropic modes.
\section{Linearized Navier-Stokes equation}
\begin{figure*}
\begin{center}
\subfigure[\ d=3]{
\includegraphics[scale=0.48]{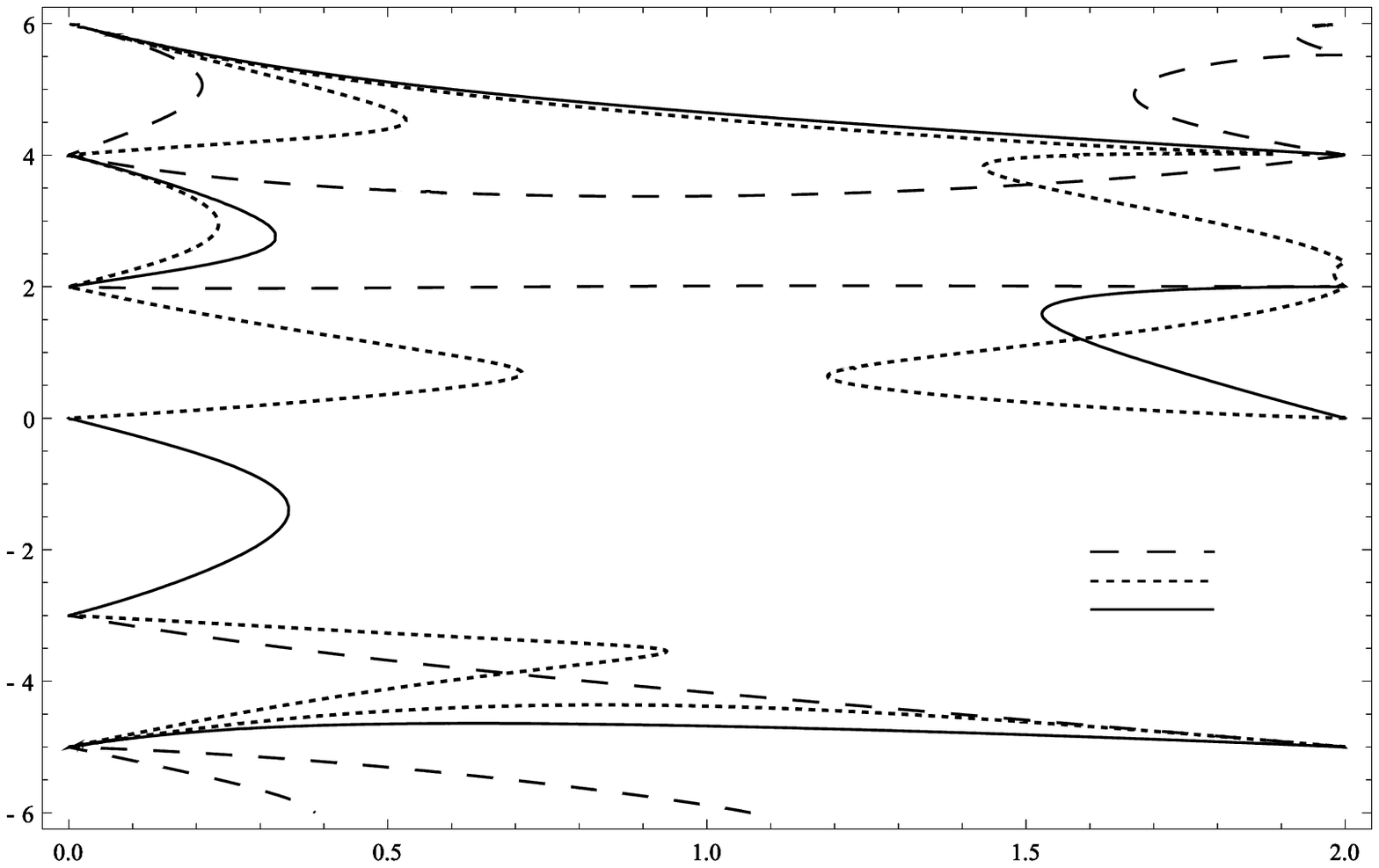}
\put(-23,49){\tiny{$l=4$}} \put(-23,44){\tiny{$l=2$}}
\put(-23,39){\tiny{$l=0$}}} \subfigure[\ d=2]{
\includegraphics[scale=0.48]{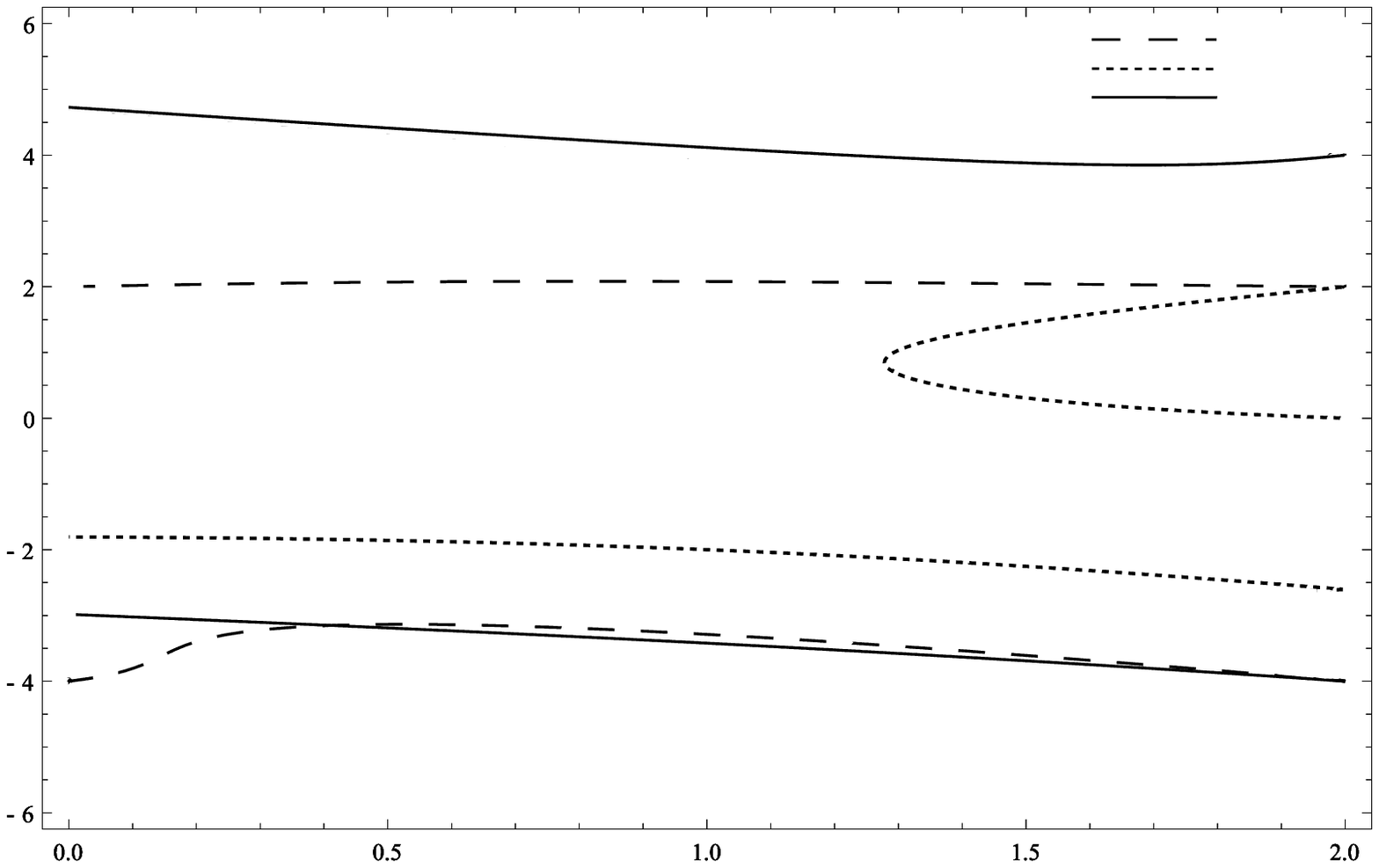}
\put(-23,132){\tiny{$l=4$}} \put(-23,127){\tiny{$l=2$}}
\put(-23,122){\tiny{$l=0$}} } \caption{The Linearized
Navier-Stokes equation exponents for sectors $l=0$, $l=2$ and
$l=4$ (the legend applies to both figures)
 at three and two dimensions.
In (a) the $l=4$ curves run slightly below and above the curves
$z=-3-\xi$ and $z=2$, respectively. Other than leading exponents
are also displayed.} \label{LNSL0toL4}
\end{center}
\end{figure*}
Setting $a=-1$ in eq. (\ref{equation}) yields the Linearized
Navier-Stokes equation (see e.g. \cite{frisch,frisch.paper,landau}). The equation may be considered as zeroth order perturbation theory of the full Navier-Stokes turbulence problem, from which one can at least in principle proceed to higher orders in perturbation theory. It will also serve as a stability problem where the background flow is determined by the Kraichnan ensemble instead of a solution to the Navier-Stokes equation (see chapter III of \cite{landau}). Not
much is known of this case, except for the perturbative results in
\cite{paolo,jurc1}. The eq. (\ref{1overgamma}) becomes
%
\eq{&&\frac{1}{\gamma_{-1}(z)} = d (d+2+\xi)
 \Gamma \left(1+ \xi/2 \right)\Gamma \left( 1+d/2 \right)
\nonumber
\\ &&- 2 p_{\footnotesize{-1}}(z) \frac{\Gamma \left( -z/2 \right)  \Gamma \left( \frac{d+z+\xi}{2} \right)
\Gamma \left( \frac{4+d-\xi}{2} \right)}{\Gamma \left(
\frac{2+d+z}{2} \right) \Gamma \left( \frac{4-z-\xi}{2} \right)
\nonumber},}
with
\eq{p_{-1} (z) &&=(z+\xi-2)  \left( d^2 (z + \xi) - (z + \xi)^2
 + d (z-1) z + d (\xi-1) \xi \right) - 4 (d+1) \xi z.}
We choose to save space by not writing down explicitly the
determinant for the anisotropic sectors. The expression may be
reproduced by using the results of appendix \ref{tauappendix}. We
will also refrain from explicitly writing down expressions for the
correlation functions, as it turns out that whichever sector has
the leading exponents varies quite a bit with different values of
$\xi$.
\subsection{d=3 with zero charge forcing}
The contour bound is, as usual, $-3-\xi < \mathcal R e (z) < 0$
and again one observes a cancelation of the corresponding pole.
Inspecting Fig. (\ref{LNSL0toL4}) one observes quite wild behavior
of the various scaling exponents at a first few sectors. A notable
similarity to the three dimensional MHD case ($a=1$) are the
exponents starting at $0$ and $-3$ and joining at $\xi \approx
0,35$. However in the LNS case one also sees similar behavior near
$\xi = 2$. Indeed one is tempted to assume the existence of a
steady state only for $\xi$ near zero and two. The same conclusion
could be drawn for the anisotropic sectors as well. We will
further discuss this at the end of the paper. We will be satisfied
with only reporting the scaling behaviors as the procedure for
finding them is close to above cases. Assuming the steady state
exists for $\xi$ close enough to zero and two, we conclude that
for $\xi$ near zero, the small and large scale are dominated by
the isotropic exponents starting at $0$ and $-3$, respectively.
For $\xi$ near $2$, one instead observes $l=2$ dominance at small
scales and $l=4$ dominance at large scales. We have deliberately
neglected the nonzero charge forcing, as that would only bring
about the familiar nonanomalous $-1-\xi$ scaling at large scales.
\subsection{$d=2$ with zero charge forcing}
The behavior of the scaling exponents are much nicer, as can be
seen by looking at Fig. (\ref{LNSL0toL4}). For $0 \le \xi \lesssim
1,3$, we see the small scales dominated by the $l=4$ anisotropic
sector, and the large scale by the $l=2$ sector. For other values
of $\xi$ the $l=2$ anisotropic sector dominates the small scales
as well. The $l>4$ anisotropic exponents are all subleading with
respect to the ones in the figure, and indeed respect the usual
hierarchy of exponents \cite{paolo}. In any case, the isotropic
exponent is subleading.
\section{The effect of varying the parameter $a$}
\begin{figure*}
\begin{center}
\subfigure[]{
\includegraphics[scale=0.48]{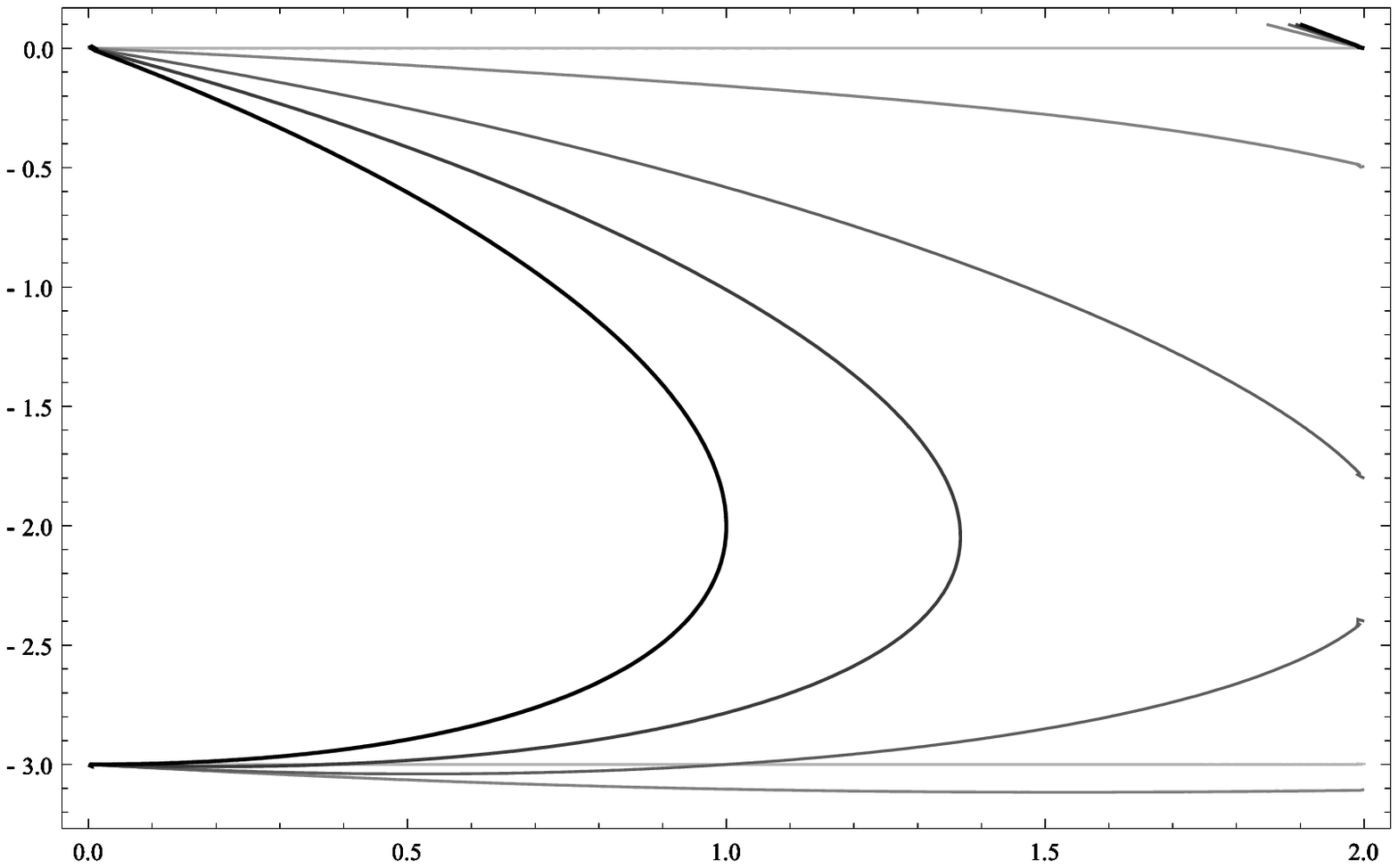}
} \subfigure[]{
\includegraphics[scale=0.48]{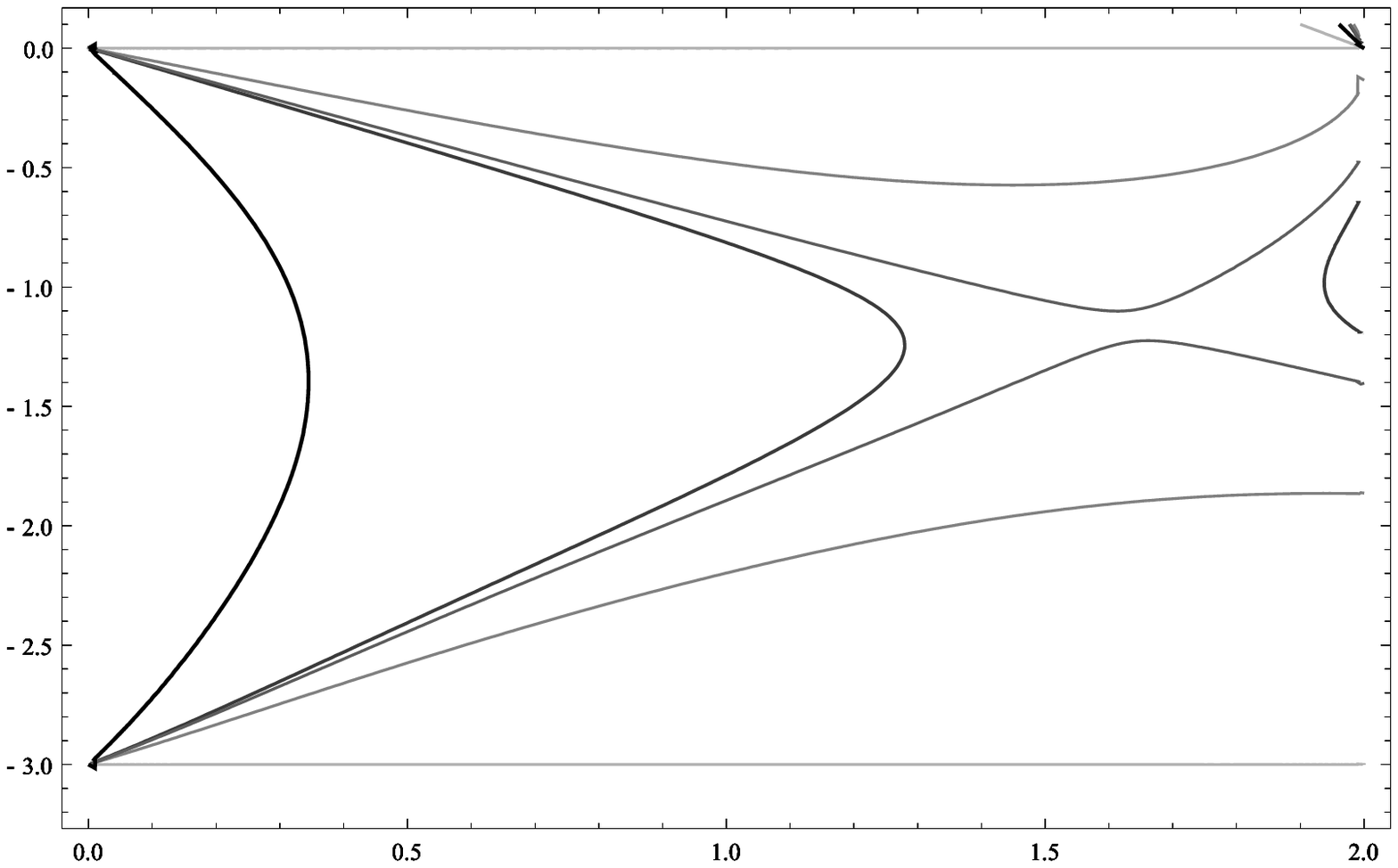}
} \caption{The leading isotropic exponents as $a$ is varied from
$1$ to $0$ (a) and from $-1$ to $0$ (b) in three dimensions. The
darkest curves correspond to $a=1$ and $a=-1$.} \label{a0toa1}
\end{center}
\end{figure*}
It is useful to discuss also other values of $a$ beside the
discrete values $a=1, 0, -1$. More specifically, looking at Fig.
(\ref{a0toa1}) we see how the closed contour determining the
leading scaling exponents is deformed as $a$ varies from $a=1$ and
$a=-1$ to $0$. Both end up as curves $z=0$ and $z=-3$. Also, as we
know that when $a=1$ the steady state exists for $\xi <1$ in the
isotropic sector \cite{vergassola} (and for all $\xi$ in the
anisotropic sectors \cite{arad}), it now seems even more
reasonable to expect the steady state to exist for all $\xi$ in
the $a=0$ case.
\section{Conclusion}
The purpose of the present paper was to present an exact solution
for the two point function of the so-called $a$ -model in the
small and large scaling regimes, which incorporates the
magnetohydrodynamic equations, the linear pressure model and the
linearized Navier-Stokes equations. The phenomena of anomalous
scaling and anisotropy dominance was investigated in each model
with emphasis placed in the zero charge forcing concentrated at a
finite wavenumber $\sim 1/L$ as in \cite{fouxon}. Below we briefly
summarize the findings in each model.\\ \\
For the magnetohydrodynamic equations with $a=1$ the leading
scaling behavior was observed to be anomalous and isotropic at
both small and large scales in three dimensions for the zero
charge forcing, in accordance with previous small scale results
\cite{arad,antonovlanottemazzino,lanotte2}. In two dimensions
with nonzero charge forcing one observes anomalous and
anisotropic behavior at small scales, while the large
scales are dominated by logarithmic behavior. The mechanism
of the small scale anisotropy dominance is strikingly similar
to the passive scalar \emph{large} scale anisotropy dominance,
except that in the MHD case the phenomenon results from the
\emph{nonzero} charge forcing. The zero charge forcing case
in two dimensions is in agreement with the results in
\cite{vergassola}.\\ \\
For the linear pressure model with $a=0$ and zero charge
we recovered the small scale exponents of \cite{adz}.
The small scale behavior is now dominated by the isotropic
and canonical scaling exponent $z=2-\xi$ (neglecting the
constant mode by considering the structure function).
The large scale behavior was seen to be dominated by a
curious isotropic zero mode $z=-d$, although the
\emph{trace} of the structure function exhibits anomalous
and anisotropic behavior at large scales. The nonzero
charge forcing simply renders the large scale behavior
canonical. The existence of the steady state is
nevertheless controversial in two dimensions and requires
further study.\\ \\
The linearized Navier-Stokes equations corresponding to $a=-1$
seem to be the most interesting of the models considered, even
more so because it is also the least well known. There still
remains the question of the existence of the steady state, without
which one cannot claim to have completely solved the problem. One
may however conjecture it's existence at least for small enough
$\xi$ (at least in the isotropic sector), in which case the small
and large scales are dominated by the isotropic anomalous scaling
exponents in three dimensions. In two dimensions, the small scale
exponents coincide with the somewhat rough numerical estimates of
\cite{yoshida}, the difference now being the absence of the
scaling $\propto r^{-\xi}$ due to the forcing. Indeed, it was
observed that both the small and large scales were dominated by
anomalous \emph{anisotropic} scaling
exponents.\\ \\
Although the linear equations above with the somewhat crude Kraichnan model are certainly some distance from the real problem of turbulence, similar scaling behavior has been observed in real and numerical simulations (see e.g. \cite{experimental,experimental2} and references therein), namely implying that the scaling exponents in each anisotropic sector are universal as outlined above. Probably the closest case to the real Navier-Stokes turbulence is the linearized Navier-Stokes equation. The equation arises usually as one tries to verify the stability of a given stationary flow by decomposing the velocity field as $v+u$, where $v$ is the stationary, time independent term and $u$ is a small perturbation \cite{landau}. If one can show that $u$ decays in time, the velocity field $v$ is indeed a laminar, stable flow. In our case $v$ is determined by the Kraichnan model and we are now concerned with the stability of the statistical steady state. It has been pointed out in \cite{jurc1} that in such a case one might be able to show that higher order perturbative terms are irrelevant in the sense of the renormalization group, thus implying that the steady state is in fact in the same universality class as the full NS turbulence. This would mean that the anomalous scaling exponent of the linear model is equal to the NS turbulence exponent. All this would of course depend on the existence of the steady state for $u$. Unfortunately it seems that such a steady state does not exist for the exponent $\xi=2/3$, which could be a sign of incompleteness of the Kraichnan model or a symptom of the general complexity of the problem of turbulence. The stability and existence problem will be studied more carefully in a future paper by the present author.
\begin{acknowledgements}
The author wishes to thank P. Muratore-Ginanneschi, A. Kupiainen
and I. Fouxon for useful discussions, suggestions and help on the
matter. This work was supported by the Academy of Finland
"\emph{Centre of excellence in Analysis and Dynamics Research}"
and TEKES project n. 40289/05 "\emph{From Discrete to Continuous
models for Multiphase Flows}".
\end{acknowledgements}
\appendix
\section{Equation of motion for the pair correlation
function \label{APXeom}}
We take the Fourier transform of equation (\ref{equation}) and
rewrite it as a stochastic partial differential equation of
Stratonovich type as
\eq{ &&d \widehat u_i (\map) = \nu p^2 \widehat u_i (\map) dt -
\widehat \D_{i\mu\nu}^{p} \inte \ud^d \maq d \widehat V_\nu (\maq)
\circ \widehat u_\mu (\map-\maq)+ d \widehat F_i (\map),}
where we have dropped the $t$ -dependence and denoted
\eq{\widehat \D_{iab}^{p} = i \left( \delta_{i a}  p_b - a
\delta_{i b} p_a \right) + i (a-1)  p_i  p_a  p_b /p^2}
and defined the Stratonovich product
\eq{d \widehat V_\nu (\maq) \circ \widehat u_\mu (\map-\maq) = d
\widehat V_\nu (t,\maq) \widehat u_\mu (t+\frac{dt}{2},
\map-\maq).}
As argued in \cite{zinn-justin} by physical grounds, the symmetric
prescription $\theta (0) = 1/2$, corresponding to the Stratonovich
definition of the SPDE, is the correct way of defining the
equation. We will however use the relation $\widehat u_\mu
(t+\frac{dt}{2}, \map) = \widehat u_\mu (t, \map) + \frac{1}{2} d
\widehat u_\mu (t, \map)$ to transform the equation into a
following It\^o SPDE,
\eq{ &&d \widehat u_i (\map) = \nu p^2 \widehat u_i (\map) dt +
\frac{1}{2}\widehat \D_{i\mu\nu}^{p} \inte \ud^d \maq \widehat
D_{\nu\sigma} (\maq) \widehat \D_{\mu\rho\sigma}^{p-q} \widehat
u_\rho (\map) \nonumber
\\ &&\ldots- \widehat \D_{i\mu\nu}^{p} \inte \ud^d \maq d \widehat V_\nu
(\maq) \widehat u_\mu (\map-\maq)+ d \widehat F_i (\map),}
where we have used the relation
\eq{d \widehat V_i (t, \map) d \widehat V_j (t,\map') = \widehat
D_{ij} (\map) \delta^d (\map + \map') dt.}
The first integral on the right hand side of the It\^o SPDE can be
done explicitly, resulting in
\eq{&&-\frac{1}{2}\widehat \D_{i\mu\nu}^{p} \inte \ud^d \maq
\widehat D_{\nu\sigma} (\maq) \widehat \D_{\mu\rho\sigma}^{p-q}
\widehat u_\rho (\map)= D m_v^{-\xi} p^2 \widehat u_i (\map) +
\widetilde \lambda p^{2-\xi} \widehat u_i (\map) + \mathcal O
(m_v^+),}
where the incompressibility condition $p_i \widehat u_i (\map) =
0$ was used, and denoting
\eq{&&\widetilde \lambda = (a-1)\left(d+1+a(1-\xi)\right)
 \frac{d \pi \xi \csc (\pi \xi/2) \Gamma
(d/2) c_d}{16 \Gamma \left( \frac{d-\xi}{2}+2 \right) \Gamma
\left( \frac{d+\xi}{2}+1 \right)}.\label{lambda}}
Applying the It\^o formula to the quantity
\eq{\langle \widehat u_i (t,\map) \widehat u_j (t,\map')\rangle
\doteq \widehat G_{ij} (t,\map) \delta^d (\map+\map')}
and by assuming stationarity, one obtains the nonlocal PDE (with
obvious $\map$ dependence omitted)
\eq{&&\left[ 2 \nu - D m_v^{-\xi} \right] |\map|^2 \widehat G_{ij}
- \widetilde \lambda D_0 |\map|^{2-\xi} \widehat G_{ij}+ \widehat
\D_{i\mu\nu}^{ p} \widehat \D_{j\rho\sigma}^{- p} \inte \ud^d \maq
\widehat D_{\nu \sigma}(\maq) \widehat G_{\mu \rho}(\map-\maq) =
\widehat C_{ij}. \label{fouriereq}}
Using the $\mathrm{SO(d)}$ decomposition for $\widehat G$,
\eq{\widehat{G}_{ij} (\map) := \sum\limits_{a} B_{ij}^{a} (\hat
\map) \widehat G^a (p)}
(and similarly for $\widehat C$), dividing the equation by $p^2$,
and by taking the Mellin transform of the equation while
remembering the definition
\eq{&&\widehat{G}_{i j}^z (\map) = \int\limits_0^\infty
\frac{dw}{w} w^{d+z} \widehat{G}_{ij} (w \map)= |\map|^{-d-z}
\sum\limits_q B_{i j}^a (\hat \map) \bar g_a
(z),\label{mellinofcorr}}
and by expressing $\widehat D$ in the integrand as an inverse
Mellin transform, we finally obtain the equation
\eq{\left[2 \nu - D m_v^{-\xi}\right] \bar g_b (z)\!\!\!\!\!\! &&-
\widetilde \lambda
 D_0 \bar g_b (z-\xi) + \inte \ud z' \bar d_{m_v} (z')
 \mathrm T_{d+z',d+z-z'}^{bc} \bar g_c
(z-z')
 \nonumber \\ &&= \bar c_b (z-2), \label{A12}}
where we have defined (note the transpose in definition)
\eq{&&\sum\limits_b \mathrm T_{d+z',d+z-z'}^{cb} B_{i j}^c (\hat
\map) = |\map|^{d+z-2} \D_{i\mu \nu}^{ \map} \D_{j\rho \sigma}^{-
\map}  \inte \ud^d \maq \frac{
 P_{\nu \sigma}
(\map - \maq) B_{\mu \rho}^b(\maq)}{\left| \map - \maq
\right|^{d+z'} \left| \maq \right|^{d+z-z'}} \label{Lambda},}
with the strips of analyticity,
\eq{\mathcal R e (z)-\mathcal R e (z') &&<0 \nonumber \\
    \mathcal R e (z') &&<0 \nonumber \\
    d+ \mathcal R e (z) &&>0\label{strips1}}
such that the $9 \times 9$ matrix $\mathbf T$ is independent of
$\map$. The matrix elements $\mathrm T^{bc}$ can be determined
exactly by computing the right hand side integral, which is the
subject of the next appendix. As mentioned in sec.
\ref{steadystate}, the first poles on the right occur at $z'=0$
and $z'=\xi$, which results in the equation in the limit of
vanishing $m_v$:

\eq{2 \nu \bar g_b (z) - D_0 \widetilde \lambda \bar g_b (z-\xi)
\!\!\!\!&&+ \ \bar d_{m_v} (0) \mathrm R^{bc} \bar g_c (z)  - D_0
\mathrm T^{bc}_{d+\xi,d+z-\xi} (z) \bar g_c (z-\xi) \nonumber \\
&&= \bar c_b (z-2).}
We have defined the residue matrix

\eq{\mathrm R^{bc} = \mathcal{R} \left( \mathrm
T_{d+z',d+z-z'}^{bc} \right)|_{z'=0}\label{Rmatrix}}
and used the residue of the velocity correlation at $z'=\xi$:
\eq{ \mathcal R_{z' = \xi} \left( \bar d_{m_v} (z') \right) =
-D_0.}
\section{Incompressibility condition\label{APXincomp}}
The incompressibility condition for $u$ and $f$ amounts to
requiring that the contraction of the covariances
(\ref{covariances}) with $\map$ is zero, i.e.
\eq{|\map|^{d+z+l} p_i \widehat G_{ij}^z (\map) &&= \left( p_j
\bar g_1 + l p_j \bar g_3 + p_j \bar g_4 \right) \Phi^l (\map) +
|\map|^2\left((l-1) \bar g_2
\partial_j +  \bar g_3
\partial_j  \right) \Phi^l (\map) \nonumber \\
&&\equiv 0,\label{incompcond}}
which gives a system of equations
\eq{\bar g_1 + l \bar g_3 + \bar g_4 &= 0 \nonumber \\ (l-1) \bar
g_2 +\bar g_3 &= 0.\label{systemofequations}}
We can achieve this conveniently by defining a projection operator
\eq{\widehat{\mathbf P} = \begin{pmatrix}
 \mathbf{1} & 0 \\
 \mathbf X & 0
\end{pmatrix},\label{projector}}
where
\eq{\mathbf X = \begin{pmatrix}
 0 & -(l-1) \\
 -1 & l(l-1)
\end{pmatrix}.}
The solution to eq. (\ref{systemofequations}) (and a similar one
for the forcing) can then be written conveniently as
\eq{\mathbf{\bar g} := \begin{pmatrix}
 \mathbf{\bar h} \\  \mathbf X \cdot \mathbf{\bar h}
\end{pmatrix}  ;  \mathbf{\bar c} := \begin{pmatrix}
 \mathbf{\bar f} \\ \mathbf X \cdot \mathbf{\bar f}
\end{pmatrix} .}
We also rewrite the matrices $\mathbf R$ and $\mathbf T$ in block
form as

\eq{\mathbf R = \begin{pmatrix}
 \mathbf R_1 & \mathbf R_2\\
 \mathbf R_3 & \mathbf R_4
\end{pmatrix} ; \mathbf T_{d+\xi,d+z} =
\begin{pmatrix}
 \mathbf A & \mathbf B \\ \mathbf C & \mathbf D
\end{pmatrix}.}
Note the above definition of $\mathbf T$ with a translation $z \to
z+\xi$. $\mathbf R$ is independent of $z$. By operating with
$\mathbf{\widehat P}$ on eq. (\ref{eq2}), we obtain the equations
(after translation $z \to z+\xi$),
\eq{&&\left[2 \nu\! -\! D m_v^{-\xi}\right] \mathbf{\bar
h}(z+\xi)\! + \bar d_{m_v} (0) \left( \mathbf R_1\! + \mathbf
R_2\! \cdot\! \mathbf X \right)\! \mathbf{\bar h}(z+\xi)
-\widetilde \lambda D_0 \mathbf{\bar h}(z)- D_0 \left( \mathbf A+
\mathbf B \cdot \mathbf X \right) \mathbf{\bar h}(z) \nonumber \\
&&= \mathbf{\bar f}(z+\xi-2)\label{eq3}}
and an identical one but multiplied by $\mathbf X$ from the left.
Thus we see that we only need the upper 2 by 2 matrices from
$\mathbf T$. By using the definition eq. (\ref{Rmatrix}) and the
results for $\mathrm T^{ab}$ in appendix \ref{tauappendix}, we
obtain

\eq{\mathbf R_1+ \mathbf R_2 \cdot \mathbf  X =
-\frac{d-1}{\Gamma(d/2+1)} c_d \mathbf{1},}
%
%
%
%
%
which results in a cancellation of the remaining mass dependent
terms.
The remaining equation depends now only on the physical
diffusivity $\nu$. Solving the equation iteratively would amount
to a series expansion in powers of $\nu$ or $\nu^{-1}$, but we
shall only consider the $\nu \to 0$ limit, which produces the
solution in eq. (\ref{eq2}).
\section{Necessary components of the matrix $\mathbf{T}$ \label{tauappendix}}
Due to incompressibility, only some of the components of $\mathbf
T$ will be needed. Computing the integrals of the type in
(\ref{Lambda}) can be performed by using the result
\eq{\int \ud^d \maq \frac{\Phi^l (\widehat \maq \cdot \widehat
e)}{\left| \maq \right|^{2 \alpha} \left| \map -
\maq\right|^{2\beta} } =: \lambda_{2\alpha, 2\beta} \left| \map
\right|^{d-2(\alpha + \beta)} \Phi^l (\widehat \map \cdot \widehat
e),}
where we have denoted by $\widehat \maq \cdot \widehat e$ the
angle between $\maq$ and the $z$ -axis and defined
\eq{\lambda_{2\alpha, 2\beta} := \frac{\Gamma (d/2+l-\alpha)
 \Gamma (d/2-\beta) \Gamma (\alpha+\beta-d/2)}{\Gamma
  (\alpha) \Gamma (\beta) \Gamma (d+l-\alpha-\beta)}.}
The tensorial structure can be obtained by partial integrations
and by taking derivatives in $\map$. We will further define (note
the transpose in the definition)
\eq{\mathrm T^{ab}_{d+\xi,d+z}:=
\frac{\lambda_{l+d+z,d+\xi}}{d+\xi} \tau^{ab} (z).}
The necessary components of $\tau$ are (others do not contribute
due to the incompressibility condition):
\begin{widetext}
\eq{\tau^{11} &&= \frac{(1+a^2)(d-1)(l-z)- a^2
\xi(z+d+\xi-l)}{(l-z-\xi)}+ \frac{l (l-1)
\xi}{(l-z-\xi)(l+z+d+\xi-2)} \nonumber \\  \tau^{12} &&= \frac{
a^2 \xi}{(l-z-\xi)(l+z+d+\xi -2)} \nonumber }

\eq{\tau^{21} &&= a^2 l (l-1) \frac{l+z+d-2}{l+z+d+\xi-2} \left(
d-1+\xi \frac{z-l+d+\xi+2 }{z-l+2} \right) \nonumber \\ \tau^{22}
&&= \frac{(d-1)(l+z+d-2)}{l+z+d+\xi-2} + \frac{(l-2)\xi \left( a^2
(l-3) +2 a (z+d+1) +l-3 \right)}{(l-z-2)(l+z+d+\xi-2)} \nonumber
\\  &&+ \frac{(a-1)^2 (2-\xi) \xi (l^2-5
l+6)}{(l-z-2)(l+z+d+\xi-2)(l+z+d+\xi-4)} \nonumber }

\eq{\tau^{31} &&= \frac{2 a l \xi
(z+d+\xi-1)}{(l-z-\xi)(l+z+d+\xi-2)} + 2 a^2 l \left(
\frac{z+l(d+\xi-1) -(d+\xi)(z+\xi)}{l-z-\xi} \right. \nonumber \\
&&\left.  - \frac{(l-1)\xi (d+\xi-1)}{(l-z-\xi)(l+z+d+\xi-2)} -
\frac{(l-1)(2-\xi)(d+\xi) \xi}{(l-z-2)(l-z-\xi)(l+z+d+\xi-2)}
\right) \nonumber \\ \tau^{32} &&= 2\xi \frac{a \left( d-1+a(l-2)
+z+\xi \right)-d-1}{(l-z-\xi)(l+z+d+\xi-2)} + \frac{4 a
(l-2)(2-\xi) \xi}{(l-z-2)(l-z-\xi)(l+z+d+\xi-2)} \nonumber \\ &&+
\frac{2\xi (a-1)^2 \left( l^2-5l+6
\right)(2-\xi)}{(l-z-2)(l-z-\xi)(l+z+d+\xi-4)(l+z+d+\xi-2)}
\nonumber }

\eq{\tau^{41} &&= \frac{a^2 \left( (d+\xi)(z+\xi)-l(d+\xi-1) -z
\right)-2 a \xi}{l-z-\xi} + \nonumber
\\  &&+ \xi \frac{d+1+2 a (l-1) + a^2 (1+2d-d^2-\xi
(d-1))}{(l+z+d)(l-z-\xi)} + \frac{(a-1)^2
(d+1)(2-\xi)\xi}{(l+z+d)(l-z-\xi)(l+z+d+\xi-2)} \nonumber \\
 &&- (2-\xi) \xi \frac{(a-1)^2 \left( d^2+l(l+1) \right) + d
\left( (1+2l)(1 -2 a) + a^2 (1+3l-l^2)
\right)}{(l-z-2)(l+z+d)(l-z-\xi)(l+z+d+\xi-2)} \nonumber \\
&&+ \frac{2 (a-1)^2 (d+l) (d+1+l) \xi (\xi^2
-6\xi+8)}{(l-z-2)(l+z+d)(l-z-\xi+2)(l-z-\xi)(l+z+d+\xi-2)}
\nonumber }

\eq{\tau^{42} &&= \frac{a \xi \left( a (l+z+d) -2 (2-\xi)
\right)}{(l+z+d)(l-z-\xi)(l+z+d+\xi-2)} -\frac{(2-\xi) \xi \left(
d+3-2 a (d+1+l)+a^2 (d+3)
\right)}{(l-z-2)(l+z+d)(l-z-\xi)(l+z+d+\xi-2)} \nonumber
\\  &&+ \frac{(a-1)^2 \xi (d+3)
(8-6\xi+\xi^2)}{(l-z-2)(l+z+d)(l-z-\xi)(l-z-\xi+2)(l+z+d+\xi-2)}
\nonumber
\\  &&+ \frac{(a-1)^2 \xi
(6-5l+l^2)(8-6\xi+\xi^2)}{(l-z-2)(l+z+d)(l-z-\xi)(l-z-\xi+2)(l+z+d+\xi-4)
(l+z+d+\xi-2)}.}
\end{widetext}
\section{The matrix $\mathbf{\widehat P^{\text T} K}$\label{Kappendix}}
We defined the matrix $\mathbf K$ as \eq{\text{K}^{ab} B_{ij}^{b}
(\hat \mar) = \inte \ud^d \map e^{i \map \cdot \mar}
\frac{B_{ij}^{a,l} (\hat \map)}{|\map|^{d+z}},}
where the elements are obtained by direct computation.
Multiplication with the transpose of the projector
(\ref{projector}) yields
\eq{ \mathbf{P}^{\text T} \mathbf{K} = \imath 2^{-z} \frac{\Gamma
\left(\frac{l-z}{2}\right)} {\Gamma \left(\frac{d+l+z}{2} \right)}
\kappa ,\label{PTK}}
where $\kappa$ is now a $2 \times 4$ matrix,
\begin{widetext}
\eq{\kappa = \begin{pmatrix}
  1-\frac{1}{z+d+l} & \frac{-1}{(z+2-l)(z+d+l)} & \frac{-1}
  {z+d+l} & -\frac{z-l}{z+d+l}\\
  \frac{l(l-1)}{z+d+l}& \frac{(z+d)^2-l}{(z+d+l)(z+2-l)} &
  z+d-1 + \frac{l (l-1)}{z+d+l} & 2 (z-l) (z+d+l-2) +
  \frac{l(l-1)}{z+d+l}
\end{pmatrix}.}
\end{widetext}

\bibliography{bibsit}

\begin{thebibliography}{10}

\bibitem{paolo}
L.~Ts. Adzhemyan, N.~V. Antonov, A.~Mazzino, P.~Muratore-Ginanneschi, and A.~V.
  Runov.
\newblock Pressure and intermittency in passive vector turbulence.
\newblock {\em EPL (Europhysics Letters)}, 55(6):801--806, 2001.

\bibitem{adz}
L.~Ts. Adzhemyan, N.~V. Antonov, and A.~V. Runov.
\newblock Anomalous scaling, nonlocality, and anisotropy in a model of the
  passively advected vector field.
\newblock {\em Phys. Rev. E}, 64(4):046310, Sep 2001.

\bibitem{angheluta}
Luiza Angheluta, Roberto Benzi, Luca Biferale, Itamar Procaccia, and Federico
  Toschi.
\newblock Anomalous scaling exponents in nonlinear models of turbulence.
\newblock {\em Physical Review Letters}, 97(16):160601, 2006.

\bibitem{jurc1}
N.~V. Antonov, Michal Hnatich, Juha Honkonen, and Marian Jur\ifmmode
  \check{c}\else \v{c}\fi{}i\ifmmode~\check{s}\else \v{s}\fi{}in.
\newblock Turbulence with pressure: Anomalous scaling of a passive vector
  field.
\newblock {\em Phys. Rev. E}, 68(4):046306, Oct 2003.

\bibitem{antonovlanottemazzino}
N.~V. Antonov, A.~Lanotte, and A.~Mazzino.
\newblock Persistence of small-scale anisotropies and anomalous scaling in a
  model of magnetohydrodynamics turbulence.
\newblock {\em Phys. Rev. E}, 61(6):6586--6605, Jun 2000.

\bibitem{arad}
I.~Arad, L.~Biferale, and I.~Procaccia.
\newblock Nonperturbative spectrum of anomalous scaling exponents in the
  anisotropic sectors of passively advected magnetic fields.
\newblock {\em Phys. Rev. E}, 61(3):2654--2662, Mar 2000.

\bibitem{experimental}
Itai Arad, Luca Biferale, Irene Mazzitelli, and Itamar Procaccia.
\newblock Disentangling scaling properties in anisotropic and inhomogeneous
  turbulence.
\newblock {\em Phys. Rev. Lett.}, 82(25):5040--5043, Jun 1999.

\bibitem{experimental2}
Itai Arad, Brindesh Dhruva, Susan Kurien, Victor~S. L'vov, Itamar Procaccia,
  and K.~R. Sreenivasan.
\newblock Extraction of anisotropic contributions in turbulent flows.
\newblock {\em Phys. Rev. Lett.}, 81(24):5330--5333, Dec 1998.

\bibitem{arad2}
Itai Arad, Victor~S. L\char39{}vov, and Itamar Procaccia.
\newblock Correlation functions in isotropic and anisotropic turbulence: The
  role of the symmetry group.
\newblock {\em Phys. Rev. E}, 59(6):6753--6765, Jun 1999.

\bibitem{arponen}
H.~Arponen and P.~Horvai.
\newblock Dynamo effect in the kraichnan magnetohydrodynamic turbulence.
\newblock {\em J. Stat. Phys.}, 129(2):205--239, Oct 2007.

\bibitem{benzi}
R.~Benzi, L.~Biferale, and F.~Toschi.
\newblock Universality in passively advected hydrodynamic fields: the case of a
  passive vector with pressure.
\newblock {\em The European Physical Journal B}, 24:125, 2001.

\bibitem{bernard}
D.~Bernard, K.~Gaw\ifmmode~\mbox{\c{e}}\else \c{e}\fi{}dzki, and A.~Kupiainen.
\newblock Slow modes in passive advection.
\newblock {\em J. Stat. Phys.}, 90(3):519--569, Feb 1998.

\bibitem{mellin}
J.~Bertrand, P.~Bertrand, and J.~Ovarlez.
\newblock {\em ÃÂThe Mellin Transform.ÃÂ The Transforms and
  Applications Handbook: Second Edition.}
\newblock CRC Press LLC, 2000.

\bibitem{celani}
A.~Celani and A.~Seminara.
\newblock Large-scale anisotropy in scalar turbulence.
\newblock {\em Physical Review Letters}, 96(18):184501, 2006.

\bibitem{fouxon}
G.~Falkovich and A.~Fouxon.
\newblock Anomalous scaling of a passive scalar in turbulence and in
  equilibrium.
\newblock {\em Physical Review Letters}, 94(21):214502, 2005.

\bibitem{falkovich}
G.~Falkovich, K.~Gaw\ifmmode~\mbox{\c{e}}\else \c{e}\fi{}dzki, and
  M.~Vergassola.
\newblock Particles and fields in fluid turbulence.
\newblock {\em Rev. Mod. Phys.}, 73(4):913--975, Nov 2001.

\bibitem{frisch}
U.~Frisch.
\newblock {\em Turbulence: The Legacy of A. N. Kolmogorov}.
\newblock Cambridge University Press, 1995.

\bibitem{frisch.paper}
U.~Frisch, Z.~S. She, and P.~L. Sulem.
\newblock Large-scale flow driven by the anisotropic kinetic alpha effect.
\newblock {\em Physica D: Nonlinear Phenomena}, 28(3):382--392, 1987.

\bibitem{gawedzkikupiainen}
K.~Gaw\ifmmode~\mbox{\c{e}}\else \c{e}\fi{}dzki and A.~Kupiainen.
\newblock Anomalous scaling of the passive scalar.
\newblock {\em Phys. Rev. Lett.}, 75(21):3834--3837, Nov 1995.

\bibitem{ville}
V.~Hakulinen.
\newblock Passive advection and the degenerate elliptic operators $m_n$.
\newblock {\em Communications in Mathematical Physics}, 235(1):1, 2003.

\bibitem{jurc2}
M.~Hnatich, J.~Honkonen, M.~Jurcisin, A.~Mazzino, and S.~Sprinc.
\newblock Anomalous scaling of passively advected magnetic field in the
  presence of strong anisotropy.
\newblock {\em Physical Review E (Statistical, Nonlinear, and Soft Matter
  Physics)}, 71(6):066312, 2005.

\bibitem{kraichnan}
Robert~H. Kraichnan.
\newblock Anomalous scaling of a randomly advected passive scalar.
\newblock {\em Phys. Rev. Lett.}, 72(7):1016--1019, Feb 1994.

\bibitem{paolo.antti}
A.~Kupiainen and P.~Muratore-Ginanneschi.
\newblock Scaling, renormalization and statistical conservation laws in the
  kraichnan model of turbulent advection.
\newblock {\em J. Stat. Phys.}, 126(3):669--724, Feb 2007.

\bibitem{landau}
L.~D. Landau.
\newblock {\em Fluid Mechanics, 2nd. edition, Volume 6}.
\newblock Elsevier, 1987.

\bibitem{lanotte2}
Alessandra Lanotte and Andrea Mazzino.
\newblock Anisotropic nonperturbative zero modes for passively advected
  magnetic fields.
\newblock {\em Phys. Rev. E}, 60(4):R3483--R3486, Oct 1999.

\bibitem{vergassola}
M.~Vergassola.
\newblock Anomalous scaling for passively advected magnetic fields.
\newblock {\em Phys. Rev. E}, 53(4):R3021--R3024, Apr 1996.

\bibitem{yoshida}
K.~Yoshida and Y.~Kaneda.
\newblock Anomalous scaling of anisotropy of second-order moments in a model of
  a randomly advected solenoidal vector field.
\newblock {\em Phys. Rev. E}, 63(1):016308, Dec 2000.

\bibitem{zinn-justin}
J.~Zinn-Justin.
\newblock {\em Quantum Field Theory and Critical Phenomena, 3rd ed.}
\newblock Oxford University Press, 1996.

\end{thebibliography}
\bibliographystyle{plain}

\end{document}